%% file: 00_main.tex
\documentclass[sigconf]{acmart}

\AtBeginDocument{%
  \RequirePackage{hyperxmp}
}
\pdfoutput=1

\AtBeginDocument{%
  \providecommand\BibTeX{{%
    \normalfont B\kern-0.5em{\scshape i\kern-0.25em b}\kern-0.8em\TeX}}}

\copyrightyear{2025}
\acmYear{2025}
\setcopyright{acmlicensed}\acmConference[CHI '25]{CHI Conference on Human Factors in Computing Systems}{April 26-May 1, 2025}{Yokohama, Japan}
\acmBooktitle{CHI Conference on Human Factors in Computing Systems (CHI '25), April 26-May 1, 2025, Yokohama, Japan}
\acmDOI{10.1145/3706598.3713166}
\acmISBN{979-8-4007-1394-1/25/04}

\usepackage{xcolor}
\usepackage{fontawesome}
\usepackage{soul}
\usepackage{booktabs} 
\usepackage{longtable} 
\usepackage{array}

\newcommand{\revision}[1]{{\textcolor{black}{#1}}}
\begin{document}

%%
%% The "title" command has an optional parameter,
%% allowing the author to define a "short title" to be used in page headers.
% \title[Content-Centric Prototyping of Generative AI Applications]{\textit{``I'm Reverse Engineering the Whole Content:}'' Strategies and Setbacks in Prototyping Generative AI Applications}

% \title{Prototyping Generative AI Applications: Content-Centric Workflows, Roles, Collaboration, and Challenges}

% % Set style of source code
% \lstset{
%    language=JavaScript,
%    backgroundcolor=\color{lightgray},
%    extendedchars=true,
%    basicstyle=\footnotesize\ttfamily,
%    % basicstyle=\linespread{1.1}\footnotesize\ttfamily,
%    showstringspaces=false,
%    showspaces=false,
%    tabsize=2,
%    breaklines=true,
%    showtabs=false,
%    captionpos=b,
%    frame=tlbr, 
%    framesep=0.2cm, 
%    framerule=0pt
% }

% \usepackage{lipsum}

\newcommand\blfootnote[1]{%
  \begingroup
  \renewcommand\thefootnote{}\footnote{#1}%
  \addtocounter{footnote}{-1}%
  \endgroup
}

\definecolor{UX}{HTML}{E9D6E8}
\definecolor{AI}{HTML}{FAD9CA}
\definecolor{PM}{HTML}{B9E0D7}

\title[Prototyping Generative AI Applications]{Prototyping with Prompts: Emerging Approaches and Challenges in Generative AI Design for Collaborative Software Teams}

%%
%% The "author" command and its associated commands are used to define
%% the authors and their affiliations.
%% Of note is the shared affiliation of the first two authors, and the
%% "authornote" and "authornotemark" commands
%% used to denote shared contribution to the research.

\author{Hari Subramonyam\textsuperscript{*}}
 \affiliation{%
  \institution{Stanford University}
   \country{USA}
 }
 \email{harihars@stanford.edu}
 
 \author{Divy Thakkar\textsuperscript{*}}
 \affiliation{%
  \institution{Google}
   \country{USA}
 }
 \email{dthakkar@google.com}

\author{Andrew Ku}
 \affiliation{%
  \institution{Google}
   \country{USA}
 }
 \email{andrewku@google.com}

\author{Jürgen Dieber}
 \affiliation{%
  \institution{Stanford University}
   \country{USA}
 }
 \email{dieber@stanford.edu}

\author{Anoop Sinha}
 \affiliation{%
  \institution{Google }
   \country{USA}
 }
 \email{anoopsinha@google.com}

%%
%% By default, the full list of authors will be used in the page
%% headers. Often, this list is too long, and will overlap
%% other information printed in the page headers. This command allows
%% the author to define a more concise list
%% of authors' names for this purpose.
\renewcommand{\shortauthors}{Subramonyam, et al.}

%%
%% The abstract is a short summary of the work to be presented in the
%% article.
\begin{abstract}

Generative AI models are increasingly being integrated into human task workflows, enabling the production of expressive content across a wide range of contexts. Unlike traditional human-AI design methods, the new approach to designing generative capabilities focuses heavily on prompt engineering strategies. This shift requires a deeper understanding of how collaborative software teams establish and apply design guidelines, iteratively prototype prompts, and evaluate them to achieve specific outcomes. To explore these dynamics, we conducted design studies with 39 industry professionals, including UX designers, AI engineers, and product managers. Our findings highlight emerging practices and role shifts in AI system prototyping among multistakeholder teams. We observe various prompting and prototyping strategies, highlighting the pivotal role of to-be-generated content characteristics in enabling rapid, iterative prototyping with generative AI. By identifying associated challenges, such as the limited model interpretability and overfitting the design to specific example content, we outline considerations for generative AI prototyping. \blfootnote{\textsuperscript{*} indicates equal contribution by authors.}
\end{abstract}


%%
%% Keywords. The author(s) should pick words that accurately describe
%% the work being presented. Separate the keywords with commas.

\maketitle

\input{01_intro}
\input{02_relatedwork}

\input{03_method}
\input{04_findings}

\input{05_discussion}

\input{06_conclusion}

\bibliographystyle{ACM-Reference-Format}
\bibliography{99_refs}

\clearpage
\appendix
\input{07_appendix}

\end{document}

%% file: 01_intro.tex
\section{Introduction}

\revision{The shift from predictive} AI to generative AI, driven by advances in large language models (LLMs)~\cite{vaswani2017attention}, is transforming how we design and prototype AI applications. Unlike traditional AI development, where task-specific models are fine-tuned and trained with narrowly defined objectives, generative AI enables teams to rapidly experiment with pre-trained models with more generalized capabilities across tasks using prompt engineering. This approach empowers multidisciplinary teams --- composed of user researchers, designers, product managers, and machine learning engineers --- to quickly assess model capabilities, explore design possibilities, and iterate on solutions before committing to full-scale software development. \textit{Prompt engineering} plays a pivotal role in this new workflow, allowing teams to shape AI outputs by crafting effective input prompts. This ability to experiment quickly and interactively with AI models introduces a more \textit{dynamic} design process. User researchers, for example, can use generative AI to generate diverse user personas~\cite{ha2024clochat}, while designers can rapidly prototype user interfaces and create multiple design variations~\cite{feng2024canvil}. Product managers can explore feature ideation and market scenarios using AI-powered analysis~\cite{databricks}, and machine learning engineers can generate code snippets or debug algorithms by interacting with AI models in real time~\cite{tabnine}. 

\begin{figure*}[t!]
  \centering
\includegraphics[width=1\textwidth]{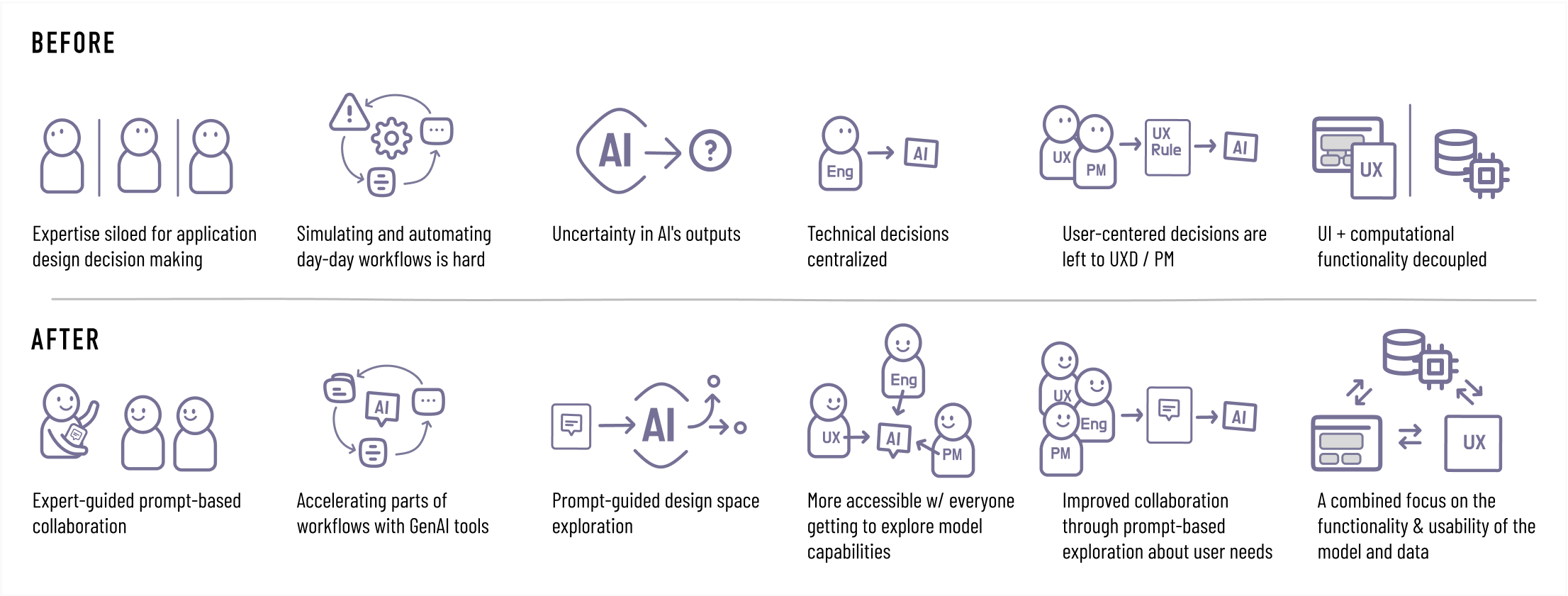}
  \caption{Figure 1. Change in Collaborative Prototyping Process with Generative AI. \revision{\textbf{Before}: Expertise (UX, PM, SWE) was siloed for application design decisions, simulating and automating workflows was difficult, uncertainty in AI outputs hindered progress, technical decisions were centralized, user-centered decisions relied heavily on UX/PM, and user interfaces were decoupled from computational functionality during prototyping. \textbf{After}: Generative AI enables expert-guided prompt-based collaboration, accelerates parts of workflows, allows prompt-guided design space exploration, makes technical and design decisions accessible to all roles, fosters collaboration around user needs, and integrates UI with computational functionality during prototyping.}
}
  \label{fig:workflows}
\end{figure*}

However, the adoption of generative AI also introduces new challenges to design workflows~\cite{morris2023design, jiang2022promptmaker}. While there is more flexibility in prototyping model outputs, there is a reduced degree of freedom in shaping the underlying AI systems, as most teams must work within the constraints of pre-trained foundational models~\cite{bommasani2021opportunities, radiya2020fine}. For designers prototyping with pre-trained models, they need to develop a designerly understanding of the pre-trained model~\cite{liao2023designerly}, find effective ways to support user interactions by minimizing the gulfs of execution, evaluation, and envisioning~\cite{subramonyam2023bridging, hutchins1985direct}, and build correspondence between interfaces, prompts, and underlying generative AI capabilities~\cite{petridis2023promptinfuser}. Further prompt engineering has its own challenges for software teams and end-users~\cite{zamfirescu2023herding,zamfirescu2023johnny}, and strategies for effectively and collaboratively designing prompts are largely lacking. 

While new tools for prompt engineering and prototyping, such as AI Studio~\cite{google_aistudio}, ChainForge~\cite{arawjo2023chainforge}, and Canvil~\cite{feng2024canvil} are being developed, we lack an understanding of how to prototype generative AI applications with them effectively. Specifically, in collaborative software teams, this new approach to design prototyping can be a complex process. The varying degree of freedom and the critical role of prompt engineering necessitates a collaborative approach to these tasks to ensure the prompt instructions and model fine-tuning are aligned with the user interfaces, task workflows, and contexts. Furthermore, the process of iteratively designing and evaluating choices during prompt prototyping remains unclear. Given these shifts in both the nature of AI design and the collaborative processes involved in prototyping, this paper explores how prompt engineering can be effectively used as a tool for rapid iteration and experimentation in generative AI application design. Specifically, we investigate the following questions:

\begin{enumerate}

\item \textit{How do teams approach the iterative prompt-based prototyping and evaluation of generative AI applications?} 
\item \textit{How do different roles within collaborative teams contribute to the prototyping process of generative AI features?} 
\item \textit{What unique challenges arise in the prototyping of generative AI applications, and how do they differ from prior AI design approaches?} 

\end{enumerate} 

We conducted a design study with 39 industry practitioners from the United States and India across 13 sessions. In each session, a UX designer, an engineer, and a product manager collaboratively worked on a given generative AI design problem and iteratively prototyped potential solutions using the AI Studio~\cite{google_aistudio} prototyping tool.  We observed that content-- both example (ground truth) content and the generated content--played a central role in the iterative design-test-analyze prototyping process, which we call \textit{content-centric prototyping}. More specifically, we observed a range of strategies in approaching prompt prototyping from templates to agent personas,  how teams collaboratively engaged in divergent and convergent thinking of formulating instructions and curating few-shot examples, and how different roles contributed to the design and evaluation tasks. We also identified key challenges to prototyping due to the high sensitivity of generative models to prompt phrases, lack of interpretability, and other human-introduced challenges, including overfitting designs to specific content examples and potential transparency issues for end-users that may be hidden at the prompt instruction level. Based on these findings, we discuss how the emerging practice of generative AI prototyping differs from prior AI experience prototyping \revision{(illustrated in Figure 1)}, shedding light on new considerations for collaborative teams working on generative AI applications.

%% file: 02_relatedwork.tex
\section{Related Work}

The shift from designing applications for task-specific ML models to generative AI models introduces new layers of complexity and flexibility to the human-centered AI design process~\cite{zamfirescu2023herding}. Task-specific models are traditionally engineered through a deliberate process of defining data and model behavior, and designers have some level of control in aligning those specifications for target user needs and interactions that are direct and predictable~\cite{subramonyam2021towards}. In contrast, generative AI, with its broader scope for content creation and autonomous decision-making, demands a more dynamic iterative design process with inherent uncertainties in design decision-making. More specifically, when prototyping generative AI applications, software teams have to balance designing in ways that harness the model's creative potential while contending with the reduced determinacy in how these models interpret and respond to user inputs, ultimately impacting the predictability of user interactions and the fidelity of generated content to user expectations~\cite{subramonyam2023bridging}. Here, we characterize this shift in design prototyping and identify knowledge gaps by integrating insights from previous studies on AI prototyping, prompt engineering in generative AI, and the roles of expertise and collaboration in implementing human-centered AI strategies.

\subsection{Prototyping AI Experiences}
In recent years, work across academia and industry has formulated design guidelines for task-specific ML models including designing interactions~\cite{heer2019agency, amershi2019guidelines, PAIR}, explainability~\cite{wang2019designing}, privacy~\cite{jobin2019global, hagendorff2019ethics}, transparency~\cite{liao2020questioning,hong2020human,bhatt2020explainable}, etc. To operationalize these guidelines, prior research has considered how software practitioners work with AI as a design material~\cite{vallgaarda2007computational, yang2018machine, liao2023designerly}, and has developed processes~\cite{subramonyam2021towards} and tools~\cite{subramonyam2021protoai,moore2023failurenotes} to prototype application experiences with AI models. Across this body of work, the basis for prototyping is that UX teams need to first understand the capabilities and limitations of ML models, including the data the model has been trained on, discerning the behaviors the model has adopted from that data, comprehending the relationships between inputs and outputs, and identify uncertainties and edge cases to the model behavior~\cite{dellermann2019future,dove2017ux,yang2020re,yang2018investigating}. Further, to facilitate this understanding, approaches such as model-informed prototyping~\cite{subramonyam2021protoai} and failure-driven design~\cite{moore2023failurenotes} support opening up the ML black box and allow designers to engage in model sensemaking using real data and model outputs that ultimately inform their design choices in prototyping. Other work has looked at deeper collaboration with engineering teams who understand ML to make design and prototyping more efficient~\cite{subramonyam2021towards,subramonyam2022solving}. However, with generative AI models, even a majority of ML engineers within individual product teams may not have the required technical understanding to facilitate designerly understanding in prototyping~\cite{zhao2023explainability}. The lack of interpretability of these generative AI models is well documented in the literature in which it is not always clear how the prompts drive the generated outputs~\cite{liu2021pretrain, sanh2022multitask}. 

Specific to prototyping, an iterative approach of design, testing, and analysis is well-established in the literature~\cite{beaudouin2009prototyping}. Prior work has emphasized leveraging data as a key element for iterating on AI system design (i.e., data probes)~\cite{helms2018design,kun2019creative, subramonyam2021towards}. That is, designers can curate data about varied users and use cases as inputs to the model and test their design choices along the way~\cite{van2018prototyping,malsattar2019designing}. However, given the size and scope of large language models~\cite{bommasani2021opportunities} and complex notions of ground truth (i.e., hard to formulate good benchmarks for correct model behavior)~\cite{subramonyam2023bridging}, it is unclear how existing data-driven approaches for AI prototyping might apply to generative AI applications. Further, in iterative prototyping, the aim is to pinpoint and rectify design breakdown~\cite{winograd1986understanding, guindon1987breakdowns, beynon2002design, rhinow2012prototypes, girardin2017user}. Given the limited interpretability of generative models, we lack an understanding of how designers might analyze the model behavior to make improvements to their design. Based on these identified constraints, our work aims to uncover how teams collectively understand the generative AI design material and how \revision{they} approach prototyping applications powered by generative models.

\subsection{Generative AI and Prompt Engineering}
Emerging work in the design of generative AI tools has put forth initial guidelines for effective prompt crafting, user interaction, and the ethical use of AI, aiming to enhance user experience and foster responsible innovation~\cite{chen2023next, weisz2023toward, subramonyam2023bridging}. Prompt engineering, a core aspect of these guidelines, involves techniques for interacting with large language models (LLMs) by establishing conversation rules, structuring output, and directing the model towards intended outcomes~\cite{white2023prompt}. Current literature draws heavily on human cognitive processes such as the \textit{chain-of-thought} technique, which outlines reasoning steps towards a conclusion and enhances the model's learning process by illustrating how to derive answers~\cite{NEURIPS2022_9d560961, kojima2023large, lampinen2022language, LOMBROZO2006167, prystawski2023psychologicallyinformed}. Another important approach, \textit{few-shot prompting}, relies on providing models with examples of desired inputs and outputs to facilitate learning from minimal information, mirroring how humans grasp new concepts from a few examples and apply them to novel situations~\cite{brown2020language, lake2019human, lake2016building, kaplan2020scaling}.  The effectiveness of these prompting methods, including few-shot and chain-of-thought prompts, is influenced by factors such as the complexity of the prompts, relevance and ordering of examples, and the incorporation of variations like Self-Consistency Sampling, Self-Ask, and Tree of Thoughts, which further refine the interaction with LLMs~\cite{fu2023complexitybased, wang2023understanding, chen2023need, wang2023selfconsistency, Stanovich_2000, press2023measuring, yao2023tree}. In our work, we aim to understand how design and engineering teams tackle the complexities of prompt engineering. This includes investigating the methods teams use to curate examples that effectively train and guide LLMs, the strategies employed to phrase prompts to achieve specific outcomes, and how these practices impact the development and refinement of prompting interfaces.

\subsection{Roles and Collaboration in Human-AI Design}
Prior work has extensively looked at different stakeholders' expertise and their collaboration challenges in operationalizing the abovementioned human-AI design guidelines~\cite{subramonyam2022solving, yang2020re}. With task-specific ML models, each role has a distinct and \textit{well-defined} function in software production; UX roles, emphasizing human psychology and design, collaborate with end-users to establish system requirements, which are then executed by software engineers proficient in computer programming~\cite{seffah2005human, amershi2019software, subramonyam2021towards, ismail2023public}. Challenges primarily emerge due to inefficient communication and collaboration. That is, UX designers may not be involved in upstream ML modeling tasks~\cite{subramonyam2022solving,yildirim2022experienced}, or domain experts may not be included in key data-related decisions~\cite{zhang2020data, zdanowska2022study}. Addressing these challenges involves overcoming the ``symmetry of ignorance~\cite{fischer2000symmetry}'' among AI professionals and other stakeholders by bridging role-specific boundaries with the aim of establishing a shared understanding, or \textit{common ground}, to enhance collaboration and interaction~\cite{fischer2000symmetry, stalnaker2002common, thakkar2022machine}. However, with generative AI, there is less engineering effort in developing bespoke models, and the emphasis is on fine-tuning the models towards specific task objectives~\cite{bommasani2021opportunities}. With the shifting nature of engineering tasks and, more broadly, designing and generative AI-powered applications, it is unclear how the different roles and expertise interact and intersect in prototyping tasks.

%% file: 03_method.tex
\section{Method}

As outlined in the research questions, our objective was to gain a deeper understanding of collaborative prototyping strategies deployed by multistakeholder teams comprising User Experience Researchers / User Experience Designers (\colorbox{UX}{UXR}), Product Managers (\colorbox{PM}{PM}), and Software Developers (\colorbox{AI}{SWE}). To this end, we conducted multi-site design workshops \revision{(n=13)} with thirty-nine participants and observed their experiences in using a Generative AI prototyping tool (AI Studio~\cite{google_aistudio}) to design an application for marketing content creation. We selected this domain because of general exposure and familiarity with marketing content across a wide range of demographics, ensuring a diverse set of perspectives and creative inputs. Consequently, the problem itself embodies the challenge of capturing and engaging diverse audiences with compelling narratives that require both creative and strategic thinking.  

We conducted the study over a three-month period from June to August 2023. Each design session involved three stakeholders -- PMs, SDEs, and UXRs and one of the authors to facilitate the sessions. The sessions lasted 105 minutes each and were conducted over Google Meet (video conferencing software). We obtained approval for this research from the Institutional Review Board of an anonymized US university. All participants signed standard consent forms. Given the observational and experimental nature of our study, our data collection involved a combination of handwritten notes and video recordings. Each design workshop was video recorded with consent. \revision{The participants are identified using a combination of their role and group number to enhance clarity in the findings. For example, `PM1' denotes the product manager in Group 1.}

\subsection{Participant Information}
We aimed to recruit one participant from each of the three roles for each session. Recruitment was carried out in collaboration with a third-party vendor that specializes in user research participant recruitment, scheduling, and disbursing incentives. We partnered with a third-party vendor for recruitment due to practical considerations that ensured the integrity and smooth execution of the multi-site study. This vendor was selected because of their ability to efficiently reach a diverse participant pool across software organizations, which was crucial for representing the varied professional roles required in our sessions. Importantly, using a vendor also ensured that participant privacy and data security were upheld to industry standards, helping us meet ethical and compliance requirements across organizations.

We chose participants who had at least one year of relevant professional experience. The recruitment process involved direct email contact and follow-up phone calls by the vendor. In one instance, one of the scheduled participants was absent, so we rescheduled the session and compensated those who were present. Participants were based in the United States of America (USA) and India; this was chosen due to the ability to conduct studies across sites given the research team's composition. After the study, participants based in the USA were provided a USD \$75 gift card, while participants based in India were provided an INR 2100 gift voucher. These amounts were set by the vendor partner, and the authors did not influence it. Table 1 describes participant profiles.

\begin{table*}[htbp]

\centering
\begin{tabular}{@{}llll>{\raggedright\arraybackslash}p{3cm}l@{}}
\toprule
\textbf{Participant Code} & \textbf{Age Group} & \textbf{Gender} & \textbf{Years of Experience} & \textbf{Level of familiarity with Generative AI/LLM [Scale of 1-5]} & \textbf{Location} \\ 
\midrule
PM1 & 31 to 40 & M & 10 & 4 & USA \\
SWE1 & 24 to 30 & M & 5 & 4 & USA \\
UXR1 & 31 to 40 & F & 1 & 5 & USA \\
\addlinespace
PM2 & 41 to 50 & M & 5 & 4 & USA \\
SWE2 & 41 to 50 & M & 5 & 5 & USA \\
UXR2 & 31 to 40 & M & 2 & 4 & USA \\
\addlinespace
PM3 & 31 to 40 & M & 6 & 3 & USA \\
SDE3 & 41 to 50 & M & 15 & 4 & USA \\
UXR3 & 31 to 40 & F & 9 & 5 & USA \\
\addlinespace
PM4 & 24 to 30 & M & 4 & 4 & USA \\
SDE4 & 31 to 40 & M & 2 & 3 & USA \\
UXR4 & 31 to 40 & M & 4 & 3 & USA \\
\addlinespace
PM5 & 31 to 40 & M & 6 & 5 & USA \\
SDE5 & 31 to 40 & M & 10 & 5 & USA \\
UXR5 & 24 to 30 & M & 9 & 5 & USA \\
\addlinespace
PM6 & 24 to 30 & M & 1 & 5 & USA \\
SDE6 & 24 to 30 & M & 3 & 4 & USA \\
UXR6 & 31 to 40 & M & 3 & 3 & USA \\
\addlinespace
PM7 & 31 to 40 & M & 6 & 4 & India \\
SWE7 & 24 to 30 & M & 3 & 5 & India \\
UXR7 & 24 to 30 & F & 3 & 3 & India \\
\addlinespace
PM8 & 31 to 40 & M & 4 & 4 & India \\
SWE8 & 31 to 40 & M & 5 & 4 & India \\
UXR8 & 24 to 30 & M & 7 & 4 & India \\
\addlinespace
PM9 & 24 to 30 & F & 1 & 4 & India \\
SDE9 & 24 to 30 & M & 5 & 5 & India \\
UXR9 & 31 to 40 & M & 12 & 3 & India \\
\addlinespace
PM10 & 24 to 30 & M & 5 & 4 & India \\
SDE10 & 41 to 50 & M & 10 & 5 & India \\
UXR10 & 24 to 30 & F & 4 & 3 & India \\
\addlinespace
PM11 & 31 to 40 & M & 9 & 4 & India \\
SDE11 & 31 to 40 & M & 14 & 5 & India \\
UXR11 & 24 to 30 & F & 2 & 5 & India \\
\addlinespace
PM12 & 41 to 50 & M & 8 & 3 & USA \\
SDE12 & 31 to 40 & F & 7 & 5 & USA \\
UXR12 & 31 to 40 & M & 5 & 3 & USA \\
\addlinespace
PM13 & 31 to 40 & M & 4 & 3 & USA \\
SDE13 & 41 to 50 & M & 10 & 3 & USA \\
UXR13 & 41 to 50 & M & 20 & 5 & USA \\
\bottomrule
\end{tabular}
\caption{Self-Reported Participant Information. Gender: F- Female, M-Male, Level of Familiarity with Generative AI on a scale of 1-5}
\end{table*}

\subsection{Tool Description: AI Studio}

Our study used AI Studio~\cite{google_aistudio} (formerly AI Studio) to investigate the prototyping process with generative models. AI Studio is a browser-based IDE that enables rapid prototyping with generative models, utilizing a version of the large language model described in~\cite{google_gemini}.

AI Studio offers two interfaces for prompting the language model: (1) \textbf{Freeform and Chat Prompts}, which allow natural language interaction for building conversational experiences, and (2) \textbf{Structured Prompts}, where users can specify example requests and expected replies for more controlled outcomes.

Participants in our study primarily focused on structured prompts due to their higher control over the model's output, though they occasionally utilized freeform and chat prompts. The typical workflow involved crafting a prompt to achieve a specific objective, running the model by clicking the "Run" button, and reviewing the output. Participants could then iterate on their prompts by refining the wording, adding more test examples, or adjusting the temperature parameter, which ranges from 0 to 1 and controls the randomness of the generated text. This allowed participants to fine-tune the creativity of the model's responses.

\begin{figure*}[t!]
  \centering
\includegraphics[width=1\textwidth]{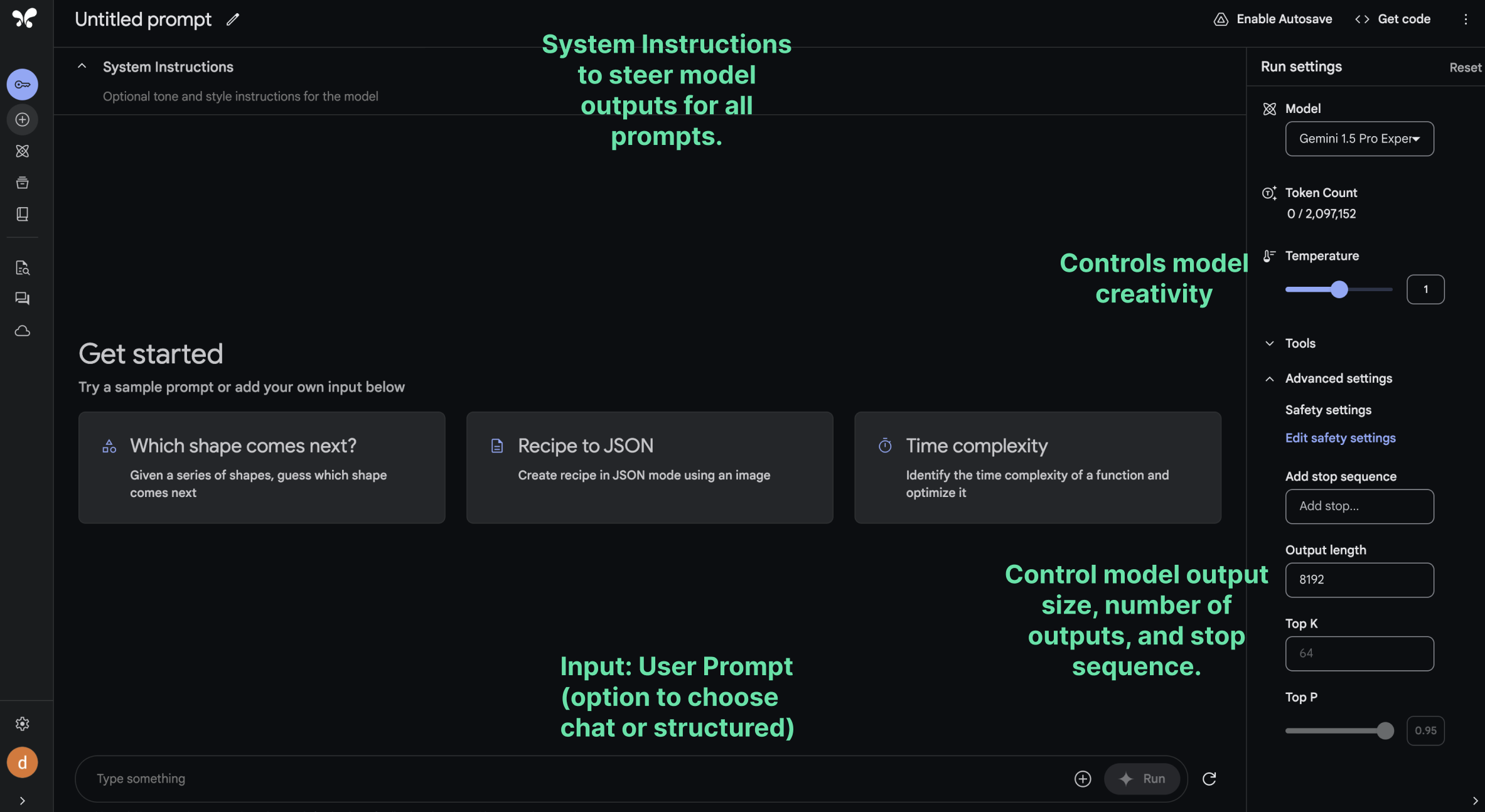}
  \caption{AI Studio provides text output only; it does not generate user interface screens or source code. However, it includes a ``Get Code'' feature, which offers a code snippet for integrating the generated prompts into external applications via the model's API. This functionality was not within the scope of our study, as we focused solely on the prototyping process.}
  \Description{}
  \label{fig:aistudio}
\end{figure*}

\subsection{Study Protocol}
We anchored our design activity on a real-world application of Generative AI in an organizational setting -- marketing content creation (detailed in Section~\ref{sec_design_activity}). We used existing professional content creation and copyright workflows based on information on the web and one of the author's own experiences as a starting point to develop instructions for the protocol. We identified \revision{introductory topics} to guide participants toward common goals. The questions corresponded to (1) Product Goals, (2) User Personas, (3) Prompt Structure, (4) Expected Outputs, and (5) Guardrails. However, these were intended as suggested thinking tools and we did not mandate their usage while running the design activity. Participants used AI Studio~\cite{google_aistudio}, to prototype the application. We should note that our intent is not to assess the tool itself. AI Studio is one of the first publicly available prototyping tools for generative AI applications. In \revision{addition} to the conversational affordances of tools such as ChatGPT, AI Studio has specific interface features to provide few-shot examples in a tabular format and also design conversational interactions. 

To establish goals and share taxonomy, workshops began with a 25-minute presentation from the lead author, who first introduced the purpose of the study and also provided a hands-on demo of the AI Studio tool with prompts and examples (unrelated to the study activity). We also provided participants with the basics of prompt engineering using the guidelines presented by White et al.~\cite{white2023prompt}. Then, before introducing the activity, participants were instructed that each of them should aim to bring in their role-specific expertise in the design activity and that this was a collaborative effort. For instance, SDEs were requested to engage in identifying relevant datasets and using AI Studio to experiment and validate against (self-defined) metrics. UXRs were motivated to validate system design choices for user needs and use AI Studio to work through user interactions and evaluate their prototype based on perceived user needs. Lastly, PMs were encouraged to define product positioning, value proposition, user personas, and metrics and assess output quality. 

At this stage, participants were requested to explore AI Studio on their own and ask the research coordinator any questions they may have. Once participants indicated familiarity with the tool, we provided a detailed description of the design activity and objectives described below. Next, participants were asked to collaboratively discuss and design a solution to the provided design brief. Participants spent between 60-70 minutes working through the activity. All participants had access to the prototyping tool via their own web browser, and participants were asked to share their screens with the tool open. In some sessions, one participant took the lead and shared their own screen for the entire session. In other cases, participants took turns in sharing the screen. Additionally, we provided participants with a shared Google Document in which they could share any prompts and contents generated through independent exploration. Throughout the session, the research coordinator asked clarifying questions or engaged in short conversations with study participants based on observations. However, we practiced caution to ensure that their work was not disrupted. The group collectively brainstormed ideas verbally. All data collected was anonymized, and videos were recorded with informed consent. 

Upon completion of the activity, we asked participants a series of reflective questions to understand their experiences of using generative AI tools for prototyping. These questions aimed to capture associations between prototyping generative AI prompts and their own expertise and the challenges they faced. In addition, we probed them about potential features for generative AI prototyping tools.

\subsubsection{Activity Description}\label{sec_design_activity}

\begin{quote}

GoodWriting Pvt Ltd is a creative writing agency that assists businesses (clients) with crafting engaging written content for their various needs. To create compelling and attractive material, the process requires a lot of creative thinking, writing, iterating, and revising of content. Clients provide a set of requirements to a marketing content creator at GoodWriting. These requirements include information such as product name, value proposition, target demographics, product description, and preferred channel of content distribution.

GoodWriting wants to use an LLM-based application to support this process. Your goal in today's session is to build a prototype for a creative content writing tool that will allow GoodWriting to do its work more effectively and efficiently.
\end{quote}

We prescribed participants to use Instagram as the channel of distribution. In the first five sessions, we provided a detailed example of a beauty product to participants and realized that participants predominantly anchored on creating a tool specific to that product rather than developing a versatile prototype that aligned with our broader design objectives. Recognizing this unintended effect of our initial approach, we incorporated a wider variety of examples, including links to two Instagram posts from both a beauty brand and a clothing brand, to emphasize the importance of generality in design objectives. This adjustment was intended to mitigate the initial oversight and align closely with our study's goals. Subsequent sessions demonstrated an improved engagement in developing a general tool. However, insights gained from the first five sessions also contribute valuable information to our overall findings. 

We provided questions to guide the design process. These questions were not prescribed and were provided to motivate the group discussion. \textit{Product Goals} We encouraged participants to think through the goals of the prototype and an eventual product vision. 
\textit{User Personas} We guided participants to reflect on the needs of the prototype user.  
\textit{Prompt Structure} We provided guiding questions to help participants identify mappings of user needs and their domain expertise to the prompt structure. 
\textit{Output Evaluation} We asked participants to define the constituents of good output based on their prompt. 
\textit{Guardrails} We asked participants to reflect on the validity and appropriateness of their tool's outputs. 
Complete details of the protocol can be found in the supplementary materials. 

\subsection{Data Analysis}
The two lead authors conducted inductive qualitative coding in Atlas.ti~\cite{atlasti} using a grounded theory approach beginning with in-vivo analysis~\cite{strauss1990basics}. Lead authors independently open-coded the same three transcripts and developed an initial code book. The resulting codebook consists of 62 high-level codes. Since using a grounded theory approach, we did not see a strong need to compute coder reliability \cite{mcdonald2019irr}. We synthesized the key emergent themes, including strategies for prototyping, prompt phrasing, design inputs, engineering inputs, friction during prototyping, model understanding, and prompt evaluation, and observed challenges and concerns such as stereotyping, biases, and generalizability. Throughout the coding process, the lead authors wrote detailed memos describing insights, observations, and emergent themes~\cite{birks2008memoing}. Memos were comprised of analyzing transcriptions as well as videos and included screenshots from the participants' design activities. After the coding was completed, the research team engaged in multiple discussion sessions to iteratively refine and converge on high-level themes and synthesize findings. 

%% file: 04_findings.tex
\section{Findings}
Across all sessions, participants had familiarity with large language models and had used services such as \revision{Gemini (formerly Bard)~\cite{bard}} and ChatGPT~\cite{chatgpt} for their work-related tasks, including summarizing documents, synthesizing data from user research, and adding representative placeholder text in interface designs, etc. Although participants were impressed by the capabilities of generative models, the majority also recognized issues with the produced text, including hallucination and low-quality output on certain technical topics. None of the participants had prior experience designing or developing generative AI applications. Given this context about participants prior exposure to generative models, we report on (1) how participants utilized their role-specific expertise to explore the model and define design objectives, (2) how they approached the given design task of prototyping a marketing content generation tool, and (3) the challenges they encountered in working with this new design material. 

\revision{We first report how collaborative teams approached formulating design goals by uniquely contributing with their domain expertise. Across sessions, teams faced a notable cold-start problem, where participants initially struggled to define concrete objectives for the generative AI application. This challenge often led to PMs taking the lead for design goal formulation. Teams collectively worked to establish success criteria, prioritizing user experience, integrating user agency into their designs, and centering safety of model outputs. Notably, teams frequently leveraged LLMs as an additional resource, using them to elicit requirements and accelerate their conceptualization process. For instance, PMs queried LLMs to propose workflows or generate design constraints, which teams critically evaluated to refine their project direction.}

\revision{Next, we highlight the unique insights into the prototyping process as teams collaborated via prompts. We observed emergent collaboration dynamics, including iteration on a baseline prompt to refine and create a reusable prompt template. Many teams worked through existing `gold examples' — high-quality input-output pairs — to align the model with their objectives. These examples were deconstructed and expanded upon to develop additional few-shot examples that illustrated specific variations in output, such as tone or audience segmentation. }

\revision{We then focus on the evaluation strategies developed by the teams at both the prompt and interface levels. Teams engaged in divergent-convergent thinking, using methods such as prompt ablations to explore how specific prompt features influenced the model's behavior and learning. Through this, teams tried to develop a predictive model of the model's generative capabilities to ensure the reliability and robustness of output quality. Teams also emphasized evaluation at the interface level, focusing on simplifying user interactions while accounting for the inherent complexity of prompting strategies that might be adopted by end-users.}

\revision{Figure 1 represents the shifts in the prototyping as teams utilize Generative AI tools across stages -- from design to development, and eventually evaluation.}

\subsection{Expertise Guided Formulation of Design Goals}
Based on the high-level design brief provided to the participants, all teams first started brainstorming on what role the generative AI should play in supporting the marketing content specialist and how to integrate the application into existing workflows. Given the broad scope of marketing content that can span diverse topics, channels, and audiences, teams initially struggled with defining concrete design objectives. Unlike conventional software design, where the focus is on defining specific requirements and functionalities in advance, AI-driven systems introduce additional challenges related to the evolving nature of content generation. While conventional conversational agent design also involves accounting for a broad range of potential tasks and user queries, generative AI systems require an even more flexible approach. The unpredictability of AI-generated content and its sensitivity to prompt engineering introduce new complexities, requiring iterative refinement and dynamic adjustments that extend beyond traditional planning and specification processes.

As designer \colorbox{UX}{UX12} commented on this \textit{cold-start problem} \textit{``I don't even know where to begin, to be honest.''} Given these complexities, we observed that, in most sessions, the PMs took charge of the conversation and emphasized understanding the end-to-end marketing process. As illustrated in the quotes below, the initial discussions focused on identifying different stakeholders and mapping the content creation workflows. 

\begin{quote}
    \colorbox{PM}{PM8}: \textit{``Before we start, we have to keep in mind that we are actually going to create this for a marketing team. So, it should have all those key points that the team will have ready with them to input into the system so they get the desired output. So that starts with who are going to be our end users for that particular ad, and which channel? And then what is the product going to be?\ldots is it like reducing the team strength of the copywriters and then having just a few of them evaluate what is coming out of our app?'' }
\end{quote}

In these discussions, the engineers and designers adopted more observant, listening roles, contributing technical insights and human-centered considerations when directed by the flow of conversation led by the PM. When discussing how the tool might impact the current roles, the designers brought \textit{human-centered values} into the system conceptualization. Across sessions, an objective was not to completely automate the creative process but to create a symbiotic relationship between the AI and human marketers. One of the designers, UX3, highlights the importance of human expertise in generating content: 

\begin{quote}
    \colorbox{PM}{PM3}: \textit{``How to make the application more appealing? Because I mean, there is value in that, right? In some sense, this someday would ultimately replace even market research. Maybe I don't know, maybe not\ldots'' }
\end{quote}

\begin{quote}
     \colorbox{UX}{UX3}: \textit{``[As a marketer,] I know my audience better than the model. So, from my perspective, I think that is my secret sauce. Like, I understand their values and emotions, and I don't want the model to search the internet for that information. Like it's my product. I just want it to take the ingredients that I'm putting together and, like, you know, blend it into an output.''}
\end{quote}

Through these initial discussions, teams began to recognize the importance of integrating human creativity with the generative capabilities of AI, deliberating aspects of control and agency across different stakeholders in the marketing pipeline. Moreover, the engineering perspective introduced the concept of ``vectors'' or key parameters (e.g., channel, product, audience, and tone) as foundational elements for marketing content generation. The engineering perspective helped focus requirements by iterating on what needed to be developed. 

\begin{quote}
     \colorbox{AI}{SWE10}: \textit{``do we also care about what imagery and color combinations and layout to be expected from the tools? Or do we let it be with the content creator?''}
\end{quote}

\subsubsection{Identifying criteria for Success}
During this initial discussion, teams also attempted to identify criteria for the success of the application. These can be grouped into \textit{content-specific} criteria and \textit{usability} criteria. From a usability perspective, teams considered metrics such as the time required to generate targeted ad campaigns, how many iterations, ease of prompting the LLM by end-users, and alignment with the current content creation workflow. For instance,  \colorbox{UX}{UX13} commented: \textit{``How many iterations do they have to take before they get something they want?''} and in session 3, the PM suggested the application should produce multiple divergent outputs to make the process more efficient. According to  \colorbox{PM}{PM3}: \textit{``So maybe success would be to generate five different options to present to their team\ldots so we save them time and help them be more creative.''} Separately, PM4 described effectiveness in terms of delight: 
\begin{quote}
     \colorbox{PM}{PM4}: ``I think there's a matter of delightful experiences as far as effectiveness goes. And that probably relates to the number of errors encountered\ldots The last point you mentioned was about the regeneration rate, which refers to the frequency of generating new prompts or repeating the same questions.''
\end{quote}

Teams discussed the idea of integrating the context of their end-users, and teams discussed providing few-shot prompts to contextualize outputs. 
\begin{quote}
 \colorbox{AI}{SWE8}: \textit{``So we'll have to define some of the shortcuts to the model -- basically just words, that what is it those words might mean, if marketing folks are talking about these words, what is it those words might mean?''}
\end{quote}

Further, teams invested a significant amount of time considering the quality of the generated output. As  \colorbox{PM}{PM4} adds: \textit{`` Perhaps there's a system to measure the effectiveness of these ideas—specifically, how many are actually utilized versus the rate at which new prompts are regenerated. That could be one way to assess it. ''}  Additionally, the importance of ensuring safety, accuracy, and adherence to political correctness in content creation emerged as pivotal benchmarks for success. The risk of producing unusable, inaccurate, or potentially offensive content highlighted the need for mechanisms to monitor and report on the robustness and appropriateness of responses generated by the models. As UX 4 mentioned: 

\begin{quote}
     \colorbox{UX}{UX4}: ``I suppose there should be a reporting mechanism for robustness, particularly concerning unusable, inaccurate, or nonsensical responses. We need to understand the models' safety features. Given that this involves advertising, considerations of political correctness also come into play, ensuring we don't inadvertently offend or overstep boundaries.''
\end{quote}

In summary, identifying success criteria highlighted the complex interplay between operational efficiency, user satisfaction, and content integrity, providing a starting point for prototyping the application. 

\subsubsection{LLM as an Expert Participant for Requirement Generation}
\begin{figure*}[t!]
  \centering
  \includegraphics[width= 0.9\textwidth]{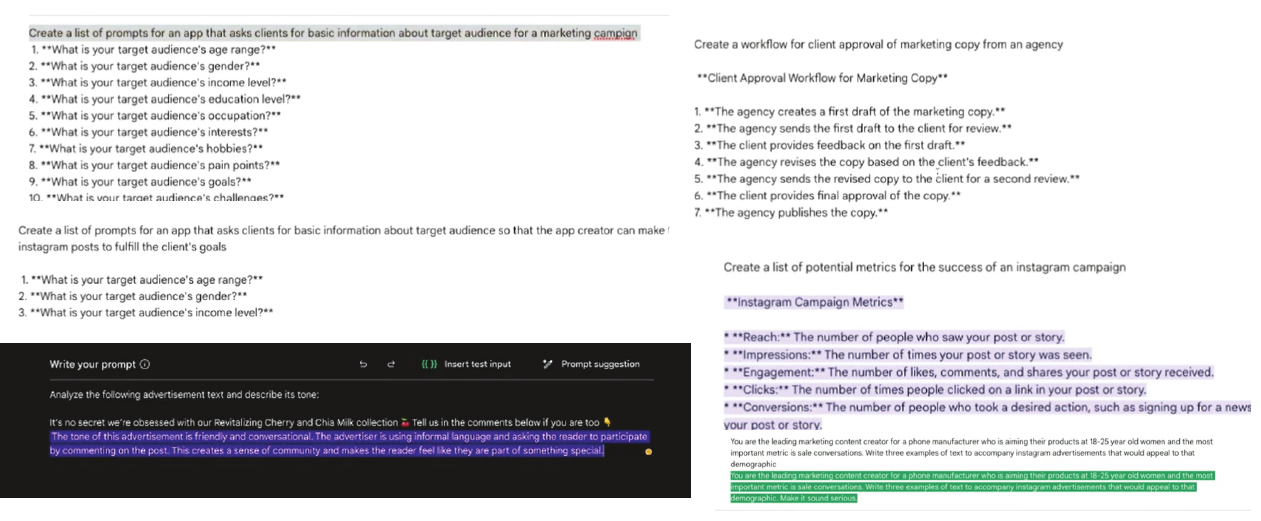}
  \caption{Examples of using LLM as a Collaboration Partner.}
  \label{fig:fuzzy}
\end{figure*}

Interestingly, in all teams, participants made use of the Gemini model as an additional expert to understand the marketing workflow and to specify design objectives (see Figure~\ref{fig:fuzzy}). For instance, PM 13 queried the LLM about devising a workflow for client content approval, leading to a simplified outline that, while not fully representative of an actual process, provided a valuable starting point for further discussion. 
\begin{quote}
     \colorbox{PM}{PM13}: \textit{``All right, what is the approval process [for the marketing content]? How are we going to decide who approves it? When is it ready [to be approved]?\ldots I asked it to come up with a workflow for client approval. So here we have the first draft the agency sends the client to provide feedback, revise, and review the final. I mean, it's so much more simple than any actual approval process with me. But at least it's given me a starting point at something to think about; I would then want to go in.'' }
\end{quote}

The preliminary workflow prompted considerations about the real-world complexity of approval procedures, highlighting the LLM's role in filling in missing expertise within software teams. Similarly, in identifying the content authoring process, manager PM4 asked Gemini to generate a potential workflow: 

\begin{quote}
     \colorbox{PM}{PM4}: \textit{`` I inquired about a marketer's workflow, and it provided a structured breakdown into research, brainstorming, writing, and editing. The research phase involves understanding the audience, the goals of the \revision{product and client}, and the competitor's landscape. Brainstorming focuses on generating headlines, outlines, and catchy phrases. Writing pertains to drafting the content, and editing involves proofreading and similar tasks. Honestly, seeing how LLMs can aid in each of these steps if we segment the application accordingly, it seems they could be quite helpful by offering specific prompts for each phase.''}
\end{quote}

This structured breakdown facilitated discussions about the overall application user experience and where the end-user agency is most needed. Further, these examples reflect a broader tendency to integrate AI into the requirement generation process, where LLMs are seen as collaborative partners with generalized expertise. However, the oversimplification of complex processes and the potential for generating outputs that lack a nuanced understanding required critical consideration amongst teams to carefully assess the generated content with their own judgment.  For instance, based on an exhaustive list of generated questions a marketer might ask their clients,  \colorbox{UX}{UX12} commented: \textit{``This is a huge overwhelming list of questions. I can help narrow those down, but then how do you want them? Presented from a UX/UI perspective, what is the most relevant information?''} Further, in several instances, based on the lack of specificity of LLM-generated requirements, teams iteratively added details to the prompts to improve the outputs.

\revision{In summary, teams encountered a cold-start problem at the start of the design stage, leading to PMs steering early discussions. Teams collaboratively identified criteria for success, including model output speed and quality, alignment with user workflows, and model output robustness. Teams leveraged GenAI tools to elicit requirements and overcome ambiguity in the design process. These findings highlight the importance of aligning team expertise and generative AI capabilities for early-stage design decisions.}

\subsection{Content-Centric Prototyping}

\textbf{Note on Terminology:} In the context of generative AI, \textit{generated content} refers to the outputs produced by the AI model in response to the input prompts. This content is typically the result of the model interpreting the prompt, applying learned patterns from training data, and generating a creative or functional response. \textit{Desired Content} refers to the expected or ideal output that the AI model aims to produce. It is often defined during the prompt engineering or prototyping stage to serve as a benchmark for what the generative AI should generate.

Using the high-level design objectives and success criteria identified through initial discussions, teams proceeded to use AI Studio~\cite{google_aistudio} to prototype prompts for the application. At a high level, we observed what is distinct about prototyping generative AI is its \textit{content-focused approach}. Unlike using a predefined set of inputs like standard forms or specific interface inputs, the goal here is to empower end-users with the freedom to contribute creative and genuine inputs and receive authentic content as outputs. Here, we synthesize our observations around prototyping strategies, the iterative process of prompt engineering and evaluation,  and interactions between prompts and envisioned user interface experiences. 

\subsubsection{Prototyping Approaches Explored by Teams} 
Across all sessions, we noticed teams took very different approaches to writing prompts. The approaches included (1) instantiating a marketing LLM agent, (2) iterating on a simple baseline prompt, (3) creating a prompt template, (4) starting with example content and authoring a prompt that would produce that content, and (5) creating few-shot examples with factors or questions identified in the initial requirements stage. In one session, the team asked the LLM to assume the role of a creative marketing specialist, focusing on how to make the AI ``think'' like a marketing expert. The focus was on how to contextually guide the LLM to produce expert-level marketing responses and how to teach the AI the nuances of marketing creativity. As Designer \colorbox{UX}{UX2} commented:\textit{ ``So the thing that I see as a limitation here is obviously, this isn't a creative marketing specialist. It's a piece of AI. So I think some of the challenges you're going to encounter are going to be around. How do you get it to think like a marketing expert?''} Hence, the focus of prompt engineering was on precisely defining and contextualizing prompts that guide the LLM to think and respond like a marketer. 

In three other sessions, participants approached the prototyping process by getting an initial assessment of how well the model performed without any specialized prompting (i.e., baseline response). This approach was in part because of not knowing how to guide the model in generating content (i.e., another case of cold-start problem). As SWE1 commented: 

\begin{quote}
    \colorbox{AI}{SWE1}:\textit{ ``It just seems like there's a lot of mental work I have to do in order to figure out what's going to happen. And discussing all of this, we're just making assumptions. But I think it really comes into play when we actually use it, see what comes out, and then go from there\ldots So I'd be curious what the base response is and then iterate from there.''}
\end{quote}

Here, iterating on a simple baseline prompt reflects a reactive approach to prompt engineering, where the initial response serves as grounds for continuous and targeted refinement. This strategy also highlights the need to interpret and debug the generated outputs in order to creatively adjust the prompts in an open-ended manner. In contrast, in two sessions, the teams initially focused on designing a structured prompt template for users to interact with based on the identified factors (tone, channel, etc.). This strategy aims to streamline the input process and more closely aligns with conventional UX prototyping that involves defining structured end-user inputs. AS \colorbox{UX}{UX3} commented: \textit{``users come in, and there are different templates for different campaigns. Then, they select a template, and they set some parameters\ldots''} However, this approach led to discussions around enabling users to effectively communicate their needs to the LLM in an unconstrained manner without restricting creativity through highly specific inputs. As \colorbox{PM}{PM7} comments: \textit{``\ldots do we prompt the user to give all the inputs or do you want to give that creative liberty for someone to input something [free form text] and then ask if something is missing?''}

In a fourth approach, which a majority of teams took, they started with some example marketing content either provided to them in the study requirements or gathered by them through the web; they attempted to reverse-engineer existing successful marketing content to create prompts that can replicate similar outcomes. As \colorbox{PM}{PM12} commented: \textit{``So then, what I was going to do is the model output will be the previously used campaign text\ldots, and I'm reverse-engineering the content.''} What this means is, by trial and error, to deconstruct the example content into constituent factors such as surprise or humor and to reconstruct prompts that elicit similar responses from the LLM. Finally, in many sessions, teams started by creating few-shot examples pertaining to each of the factors identified in the initial requirements brainstorming phase. For example, they would curate example content for Instagram vs Twitter or for different demographics such as age groups. This approach leverages the model's learning capabilities to produce content that aligns with the provided examples. As \colorbox{AI}{SWE8} highlights: \textit{``Okay. So this is what I'm trying to do. I am just trying to put some inputs and outputs\ldots I'm basically trying to define variables and teaching the parameters independently.''}

While these are initial starting points for prompt prototyping, throughout the session, teams attempted a combination of approaches. Together, these strategies highlight the prototyping process as requiring critical and creative thinking, analytical reasoning, and a collaborative effort across the different roles.

\subsubsection{Iterative Prompt Engineering}\hfill

\noindent
\textbf{Divergent-Convergent Thinking:} On detailed examination of how participants crafted the prompts, several key tasks emerged as essential to the prototyping process. Consistent with the conventional divergent-convergent UX design approach, we observed several instances of \textit{divergent thinking} where teams experimented with varying the details and scope of prompts to see how the AI responded to different contexts to find the right level of prompt abstraction. In session 13, the \colorbox{PM}{PM13} initially asked the LLM to generate an ad campaign for `` a competition the NFL (National Football League) with Gatorade (an energy drink).'' The resulting Ad included the phrase, ``Let's kick the NFL season off with some ENERGY!'' After iterating on the prompt for a bit, they replaced NFL with Tennis to assess whether the new content would reflect the updated sport or whether the prompt needed to be worded at a higher level of abstraction. In a different session, the team experimented with different tones for the generated output to determine whether and which options to give to the marketing content specialist:

\begin{quote}
    \colorbox{PM}{PM12}: \textit{``Is this tone right? Say you wanted it to be more happy, more sad, more exciting. We can give different adjectives to people to select because those can then generate inputs, very specific inputs ''}
\end{quote}

Then, to converge toward consistent and desired types of output, teams introduced \textit{constraints} into the prompts. Similarly, in many sessions, teams iterated up and down the abstraction ladder and added specificity based on the AI's feedback, gradually refining their prompts. In the convergence phase, teams discussed constraints, such as whether to restrict the use of hashtags, limit text length, add a format for the text, etc. For instance, by observing the output length of text, teams hard-coded specific parameters into the prompts (e.g., number of characters) or removed certain end-user options. A notable concern in the convergence stage was the trade-offs between safety and creativity. For instance, by observing the problematic tone of the generated output, engineer \colorbox{AI}{SWE4} pondered whether narrowing the focus of prompts by removing tone \textit{``is good from a safety perspective, I suppose\ldots''} and added \textit{`` But I wonder if it's like loss of creativity?''}  

Lastly, teams straddled between adding examples to steer the outputs vs. refining the prompt phrasing vs. adjusting model parameters when making corrections over generated output. Adding examples allowed them to specify requirements at a higher level and let the model \textit{implicitly} learn content generation needs. However, in some cases, the requirements needed to be spelled out through explicit prompting to make the output interpretable. In the convergent stage, teams frequently added explicit prompt phrases based on the model performance. As UX 3 commented:

\begin{quote}
    \colorbox{UX}{UX3}: \textit{``And it not only inferred that [tone], which is very impressive. But if we were actually going to ship this, at least for me, I would feel more comfortable knowing that we have defined parameters. For example, 'audience' has a very specific meaning, and `channel' has its own set of characteristics\ldots Shipping something that feels like a black box, like our first approach about 20 minutes ago where we were just trying to teach it to make ads—it was working [through examples], but we didn't really understand why''}
\end{quote}

Specific to setting model parameters, in several instances, teams attempted to adjust the model's \textit{temperature} (i.e., degree of randomness), that is, increase it to improve creativity, assuming the crafted prompt might be too constraining. In other cases, reducing the temperature of the output deviated from the intentions of the prompt. As PM11 commented:

\begin{quote}
    \colorbox{PM}{PM11}: \textit{``What came up was incredibly crappy, and it felt kind of patronizing. Then I said, make it sound serious, and then I changed the temperature down a little bit and asked for three different outputs. And so what we've got still came up with a bunch of exclamation points, which I could have asked for without, but it became much more the perfect framing for the Modern Woman.''}
\end{quote}

The iterative nature of prompt engineering, through adjustments in examples, temperature, and phrasing, reflects the dynamic interplay between the AI’s generative capabilities and the teams' objectives toward design objectives such as transparency and control.

\revision{In summary, teams used a collaborative approach to prototyping, simulating a marketing agent, and iterating on baseline prompts to create reusable templates. They relied on gold examples as high-quality input-output pairs and extended them into few-shot examples to refine the underlying model’s outputs. Teams used divergent-convergent thinking to explore a range of possibilities, by varying the details and scope of the prompt to assess model's sensitivity and generalization capabilities, teams then converged by building style and safety constraints into the prompt. Later, teams explored convergent strategies to across controlling model parameters (eg: temperature), providing additional few-shot examples, or by refining their prompt. These findings showcase the challenges and strategies employed in the process of collaborative prototyping. }

\textbf{Prompt Evaluation Strategies:}
We observed that prompt evaluation was rather complex and required role-specific expertise in formulating test hypotheses and dynamically defining evaluation criteria. Concretely, across all sessions, we noticed that teams formulated a range of hypotheses to gauge the AI's understanding and adaptability to inputs end-users (marketing content specialists) might provide. For instance, in the template approach, UX would change the age parameter of the prompt and assess whether the change was reflected in the new output. As \colorbox{UX}{UX3} commented: \textit{``Like if I changed the age, I am wondering if the ad will be any different''} and followed up with \textit{``It doesn't understand the concept of age.''} Following up on this, the engineer \colorbox{AI}{SWE3} suggested removing the examples and trying again with the hypothesis that the examples might be confusing to the model: \textit{``what if we don't show it any samples?\ldots I would love to have something a little bit more structured, where I can just specify parameters and ask the models to give me examples of those parameters.''} In a different session, instead of replacing the example, \colorbox{PM}{PM13} suggested interrogating the model about what it had implicitly learned from the examples: \textit{`` I have an idea. We could actually ask the model to tell us what it thinks the tone is. Let me try this''}

In addition to testing for generalizability, teams proactively engaged in adversarial testing and formulation of guardrails to identify unintended or harmful outputs. In session 6, the PM hypothesized promoting negative body image for an ad related to clothing and tested the model by varying the prompt:

\begin{quote}
    \colorbox{PM}{PM6}: \textit{``There is a possibility of unintended effects, such as promoting negative body image, which could negatively impact viewers' perception of their body. This is particularly sensitive\ldots let me try [adds to the prompt: Make the tone dark and meta, but avoid using harmful language]\ldots it might be useful to include potential warnings about how the output could be misconstrued. This wouldn't necessarily be part of the marketing campaign itself but rather an additional precaution. Is there a way we can test this separately? ''}
\end{quote}

Similarly, participants considered edge cases, such as when dealing with sensitive products like toys intended for young children, marketing efforts should be directed toward parents as well. These varied hypotheses and the corresponding model outputs were used as heuristics to determine whether the teams would consider shipping the product. In session 13, the PM expressed reservations about the readiness of the product for release, citing a narrow understanding of the model's behavior to the prompt given its limited exposure to diverse inputs and examples. 

\subsubsection{Bridging Prompt Prototyping and Interface Design}
The intersection of prompt engineering and interface design is critical to the design of generative AI applications. Across all sessions, participants explored the interplay of prompts and interactions and how best to structure user interfaces (UI) to align with and constrain the prompts. While teams did not explicitly prototype elements of the interface design, the UX designers made the connections between prompts and interface elements, such as designing a form, a questionnaire, a chat interface, a guided wizard, and even a word cloud or dropdown menu with input options. According to UX8: 

\begin{quote}
    \colorbox{UX}{UX8}: \textit{``I envision the input process like this: initially, marketing professionals would write really simple prompts. Next, they would specify the most important keywords. Lastly, they would indicate the desired platform, such as Instagram, Facebook, or newspapers. Based on these inputs, the system would then generate three outputs. That's how I see it working.''}
\end{quote}

In many sessions, teams also discussed ways to present the content to users, including multiple content outputs at once and live updates to content. For instance, in session 7, the team initially talked about an approach where the generated content would update in real time as the user manipulated the prompt parameters. The designer then mentioned difficulty tracking the input and output correspondence: 

\begin{quote}
    \colorbox{UX}{UX7}: \textit{``How do we ensure the model delivers what users are looking for on the first or perhaps second attempt, as quickly as possible, without requiring them to extensively explore options? If we continuously display live results, there's a significant chance we might lose track or not recognize if it meets the user's initial expectations, especially if they keep adjusting their inputs to experiment\ldots''}
\end{quote}

Participants were curious about the right level of information to seek from end-users, aiming to define a prompt structure that effectively captures user intent without overwhelming them with questions. UX 13 and PM 13 discuss the importance of designing a decision tree that guides users through the content generation process, suggesting a dynamic interaction model that evolves based on user input: 

\begin{quote}
    \colorbox{UX}{UX13}: \textit{``What's the maximum number of questions we can ask before it's too many? What's the user experience on the page? We need to consider the flow, how much information to provide, and the sequence of questions.''}
\end{quote}
\begin{quote}
        \colorbox{UX}{PM13}: \textit{``It seems we're creating a decision tree. Depending on the input, some of it will lead to content generation, while others will guide us through different decision-making paths based on questions and answers.''}
\end{quote}

This discussion aligns with behavioral modeling and conversational etiquette design, ensuring that the system responds fluidly to user needs while maintaining an intuitive and non-intrusive interaction. In summary, in most sessions, teams considered how users will input information, how detailed the prompts need to be, and focusing on creating a good overall experience. 

\subsection{Challenges of Prompt Sensitivity in Prototyping}
In the course of prototyping and evaluating prompts, teams grappled with complex design choices on how to effectively harness LLMs to create intuitive and efficient user interfaces that cater to the specific needs of marketing professionals. The vast array of potential questions and the overwhelming list of considerations presented significant challenges to design. Here, we present findings on three key aspects of LLM design materials that introduced friction and uncertainty in prototyping.

\subsubsection{Acquiring Predictive Models of LLMs}
In many sessions, teams encountered unexpected and surprising outputs from the LLM that both intrigued and challenged their understanding of generative AI models. For instance, in one session, the LLM added its own content around prizes for a competition when it was not something mentioned in the input prompt. As PM 12 comments: 

\begin{quote}
    \colorbox{PM}{PM12}: ``[output: It could be seen by millions of fans,] which I like. It even specified when the competition ended and what the winners would receive—details I hadn't mentioned. It also came up with the prizes on its own, which is fine by me. It's fascinating to see it mention NBA team joggers. I wonder, is that even a thing?''
\end{quote}

These instances provided valuable insights into the LLM's potential for creativity but also underscored the need for careful prompt engineering and output verification. Another significant challenge lay in breaking down the final output into sub-items and debugging outputs. Teams sought to define clear goals for the product, aiming to deliver content that precisely met the marketing team's needs. However, discrepancies in output length and content accuracy prompted reevaluation. As \colorbox{PM}{PM10} commented: \textit{``I said it should be less than 100 characters\ldots It kind of ignored your requirements. It's 174. Again, with these models, they are just so probabilistic that who knows what is actually going on\ldots''} In a similar vein, looking at the output ``imagine having skin that is so soft and smooth you can't help but touch it'' \colorbox{UX}{UX1} commented \textit{``the issue was, like, we were not able to, like, actually identify the capabilities of the tool itself properly, and how does it work?''} indicating the difficulties in aligning the LLM's outputs with specific design criteria.  Moreover, the continuous sensemaking process—where teams attempt to discern the model's behavior, highlights the challenge of reliably predicting and managing outcomes due to the model's sensitivity to specific prompt wording.

\subsubsection{Assumptions and Stereotypes In Crafting Prompts}
We observed that the process of designing prompts also brings to the forefront the nuanced challenge of navigating the team's assumptions and stereotypes. In four sessions, the LLM's response to inputs inadvertently reinforced stereotypical notions based on the way prompts were structured. For example, in discussions about targeting specific demographics, the underlying assumptions about interests and behaviors played a significant role in shaping the prompts given to the LLM. Such assumptions, while aimed at tailoring content to perceived user preferences, also highlighted the tradeoffs between personalization and the reinforcement of stereotypes. As \colorbox{PM}{PM8} noted, \textit{``When considering Tesla, the target audience we're focusing on primarily consists of males around the age group of 35. This demographic is central to our user base for this particular car category.''} In a separate session, UX1 ideated on examples for the LLM, illustrating the specificity sought in targeting the Ad yet risking oversimplification:

\begin{quote}
    \colorbox{UX}{UX1}: ``If I were to tailor a campaign specifically for females, I'd consider how to adjust the messaging differently than I would for males. For instance, a male-targeted ad might depict someone covered in mud after a hard day's work, suggesting the need for a gentle cleaner. However, I'm not sure I'd approach a female-targeted campaign in the same manner.''
\end{quote}

We can see that as teams navigate the assumptions and stereotypes inherent in crafting prompts, they are also negotiating the broader impacts of their design decisions on societal norms and values. 

\subsubsection{Overfitting Prompts to Specific Examples}
Across all sessions, teams made use of examples from existing campaigns in the prompt prototyping process. We observed that this could be a concern in the generated outputs in which the model is too closely tailored to the specific examples it has been trained on, limiting its ability to produce diverse and broadly applicable outputs. In fact, in one of the sessions, \colorbox{PM}{PM13} expressed astonishment at the model's capability to generate content from a single example: \textit{``I'm amazed that they can do this much from one example...these examples are all very similar to the first example since they're all generated from the first example''} but also noted the necessity for more comprehensive market research to ensure diversity in the generated examples. Given the lack of familiarity with prototyping generative AI applications, we noticed this tension across teams between zooming into specific use cases to refine the prompt design and the need to ensure that models can generalize across diverse scenarios.

To summarize, we find that in prototyping generative AI applications, teams constantly needed to balance the specificity of prompts with the goal of achieving broad applicability in their AI models. \revision{They used prompt ablations to analyze how specific features influenced the model’s behavior, systematically isolating and testing individual components of the prompt.} They encountered challenges like prompt sensitivity and complex evaluation needs while at the same time defining goals and metrics for success, which required careful deliberation across different roles and expertise.

%% file: 05_discussion.tex
\section{Discussion}

\subsection{Prompt Design}

\revision{We observed several unique considerations in the prototyping process. First, the iterative prompt design process resembles the process of curating `gold examples' (high-quality input-output pairs that improve model performance). Our study provides an initial insight into the curation process of such gold examples in the prototyping process. Participants engaged in strategies in `ablations', removing parts of the prompt to study model behavior and acquiring an overview of model behavior through convergent and divergent thinking. Second, the design choices that our participant teams made during the prototyping process were guided by the control knobs they had access to -- curating gold examples, prompt refinement, model parameters (eg: temperature). Our study showcases the opportunity for designers to support decision-making across such knobs.}

\begin{figure*}[t!]
  \centering
  \includegraphics[width= 0.9\textwidth]{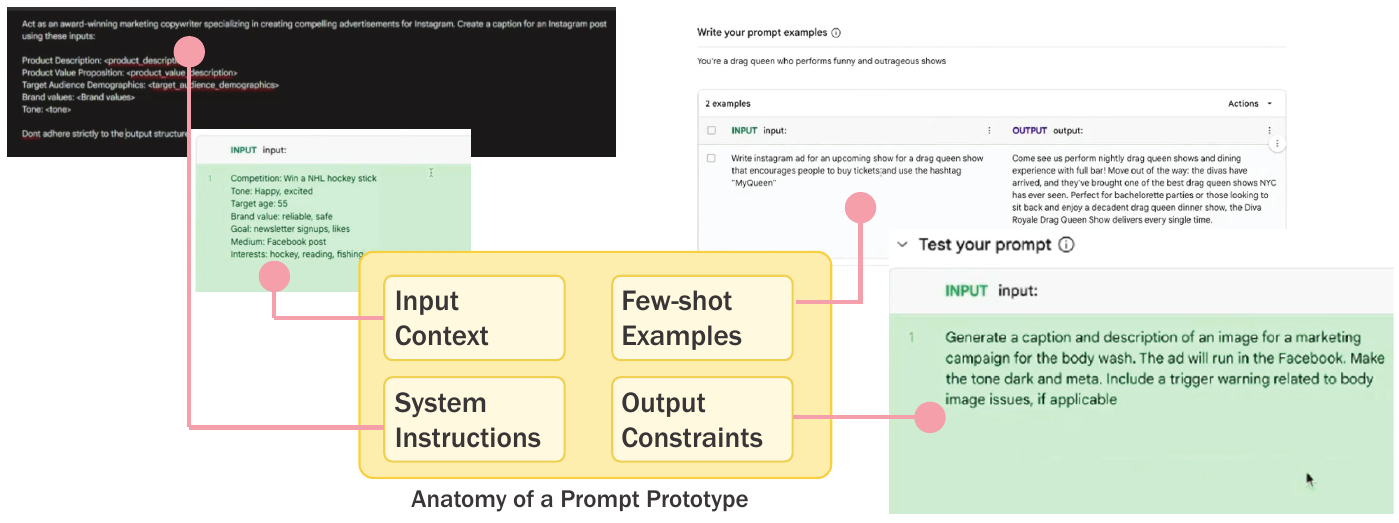}
  \caption{Four Main Components of a Prompt Prototype.}
  \label{fig:prompt}
\end{figure*}

Based on how teams authored the prompts, we identified four main sub-components of a prompt that include (1) the input context, (2) the system instructions, (3) the output constraints, and (4) few-shot examples (Figure~\ref{fig:prompt}). The input context provides the generative model with the necessary background or situational context to understand the task at hand. It encompasses specific details that the end-user might provide when using the prompt to tailor the generated response. The system instructions are commands or requests detailing what the model is expected to do and specify the task, action, or outputs desired from the AI. The output constraints provide guidelines or boundaries for the generated response. In our sessions, these constraints included stylistic aspects such as format and length of response as well as high-level guardrails, e.g., asking the model not to generate harmful content. Finally, the few-shot examples provided the model with illustrative content that implicitly indicated the desired nature of the output or behavior expected of the generative model. In addition to the `anatomy' of a prompt, our findings revealed how teams strategically approached thinking about these components and how different roles contributed to the prompt components. Designers who have expertise in crafting interfaces and designing input/output flows found themselves particularly drawn to utilizing prompts as a metaphorical `template' during the prototyping process. Conversely, certain design participants encountered challenges in understanding the concept of few-shot examples within prompts, a technique fundamentally rooted in engineering principles. Given the interconnected nature of these sub-components, and as shown in our findings-- prototyping involving refining information across the various sub-components--led to blurred lines between roles and design representations. 

This merging of roles and responsibilities has obscured individual \textbf{agency and ownership,} challenging the conventional division of tasks, such as modular design principles within project teams. Further, in identifying issues with separation of concerns, Subramonyam et al. advocate for the concept of ``leaky abstractions'' which allow for some overlap across boundaries while preserving distinct design artifacts specific to each role~\cite{subramonyam2022solving}. However, in the context of generative AI prototyping, these separations become even more \textit{indistinct}, with shared artifacts and process integration further blurring the lines between different roles. It remains an open question how this evolving dynamic will impact the definition of roles, the nature of design representations, and the methodologies employed in prototyping generative AI applications. Future research should look at how distinct roles and expertise can effectively contribute to crafting instructions for generative AI.

\subsection{Designing for Expanded Degrees of Freedom in Use}
Unlike task-specific models with focused behavior and inputs and outputs, generative AI models exhibit \textit{dynamic} capabilities and can implement ad-hoc functionality in response to prompts. Generative AI prototyping tools will increasingly need to account for the interplay in domain expertise alongside the emerging needs of collaborative software teams. 

Our findings show that teams faced challenges in deliberately crafting the specific tasks that the prompts were intended to execute. For instance, teams ideated several approaches ranging from specific prompts to set the tone of the content to broad prompts that generate the entire content given some marketing value proposition. As reported, in many sessions, teams made use of generative AI as a domain expert collaborator to generate fuzzy specifications regarding what tasks and functionalities the model should support. We foresee that executing user research to address a potentially wide array of needs (thereby utilizing the expanded degree of freedom in use that generative AI offers) could present its own set of challenges. Despite its tendency to generalize across vast amounts of information, generative AI can still serve as a valuable initial reference point for identifying and shaping prompt intentions.  Future work should focus on a \textbf{combination of these methods, enhancing the integration of generative AI in the early stages of design} and user-researcher-in-the-loop to gain insights into the diverse needs and preferences of users. 

Prototyping tools such as PromotInfuser \cite{petridis2024promptinfuser} and Claude Artefacts \cite{anthropic_artifacts_2024} provide new directions through their ability to generate dynamic user interface (UI) components. However, our study points to several opportunities to integrate emerging practices of collaborative teams in generative AI prototyping tools. We recommend prototyping tools account for agency and control over AI models (and, by extension, interaction outputs), the ability to generate synthetic data for interaction workflows, and adaptability in tools to promote collaborative teams to the forefront of their expertise.

\subsection{Evolution of Prototyping: With/for Generative AI applications}

Prototyping enables collaborative software teams to explore design ideas in a low-risk, low-investment environment, helping them clarify the scope of the final application to be built. Over the years, prototyping tools have advanced to offer high-fidelity design artifacts that effectively \textit{decouple} user experience and interactivity from underlying computing functionality. Fidelity, in this context, refers to how closely a prototype mimics the final product in terms of appearance, functionality, and interactivity~\cite{chen2024exploring}. This decoupling allowed designers to focus on user interface elements without needing to engage with the technical backend early in the process. However, with AI prototyping, the need to recouple these design artifacts with underlying AI functionality is becoming more pronounced. HCI researchers identified \revision{challenges} from the decoupled AI-first and UI-first approaches~\cite{subramonyam2022solving, yang2020re,subramonyam2021towards}, exploring ways to align AI capabilities with UX design during design. 

Our study highlights new shifts in design exploration, particularly when using generative AI tools to prototype downstream AI products. With prompt-based prototyping of generative AI, design and computation become deeply intertwined, blurring the lines between the two. Unlike traditional prototyping where design could be separated from backend functionality, generative AI requires the design process itself to function as a form of computation. Prompts serve as the mechanism through which the model generates outputs, meaning that the structure, wording, and intent of the prompts effectively become part of the computational logic. Designers are not just shaping the user experience --- they are actively configuring the AI’s behavior through the prompts they create. In this way, \textbf{the act of designing is also an act of programming, as the design artifacts directly influence the generative processes, making the computational and design aspects inseparable in this prototyping approach.}

Building on the intertwined nature of design and computation in generative AI prototyping, prior work \cite{yang2020re} has highlighted the challenges of using AI as a design material due to inherent uncertainties in AI capabilities and the complexity of its outputs. These challenges were particularly relevant to predictive AI systems, where the AI’s behavior could not be easily shaped by design teams alone. However, with generative AI, prompt-based prototyping offers new ways for collaborative teams to address these uncertainties. During our study sessions, we observed that both designers and engineers initially struggled with a ``cold-start'' problem, where each role relied on the other’s expertise or assumed that prompt prototyping fell outside their purview. To move forward, teams need to collaborate through a shared design representation, i.e., prompts.

Generative AI tools further expand the decision-making space for collaborative teams by enabling them to experiment with high-fidelity prototypes that closely resemble the final product's functionality. These tools democratize technical decisions, making choices about datasets, model providers, and even the AI's behavior more accessible to all team members. Of course, as prototyping tools such as AI Studio become central to prototyping, the concept of fidelity in design must evolve. The focus is no longer solely on simulating user interactions, but also on crafting dynamic user experiences by exploring the design space of potential end-user inputs and corresponding AI outputs. Prompting allows UX designers to iteratively test and refine AI-generated content, giving them new opportunities to influence model behavior and capabilities. Consequently, the role of UX designers is shifting from asking, “What design choices will make it easier for users to use my product?” to “What model behavior will be most useful for users?” Similarly, product managers and software engineers can rapidly prototype and test the feasibility of ideas by simulating user workflows and engaging in more active, user-centered decision-making.

\subsection{Alignment and Evaluation}
We observed key challenges in aligning interactive AI systems \cite{terry2023ai, vaithilingam2024imagining} throughout the prototyping process, particularly in how collaborative teams evaluate prototypes. Teams engage in a two-layered evaluation: they assess the validity of the model outputs and verifying correctness, while also evaluating whether end-users can comprehend and meaningfully interact with the output based on the prototype. These alignment decisions play a crucial role in shaping their prompting strategies during the prototyping process. Drawing on Terry et al.'s work on AI alignment~\cite{terry2023ai}, we find that the key components of content-centric prototyping parallel the process of achieving AI alignment. In our study, teams utilized example content to operationalize two key concepts: specification alignment, where the AI correctly interprets the user’s intent, and process alignment, where the user can influence how the AI generates the output. These strategies were employed in nuanced ways to improve the interaction between users and AI systems. We particularly observed this when participants attempted to reverse engineer prompts into specific requirements and sought to understand the underlying process behind content generation.

However, despite these benefits, our findings also revealed potential pitfalls in content-centric prototyping. When using content as a design material, teams made implicit assumptions about cultural norms, user behaviors, content formats, and concept abstractions, which may not be universally applicable. This risks excluding important user demographics and creating systems that perform well under narrow conditions but fail to address a broader spectrum of needs and contexts. Crucially, these \textbf{design choices often remain undocumented and invisible to end-users}, not clearly reflected in the user interface's affordances. For instance, the software team's predefined notions of tone or emotion embedded within a prompt may not be obvious to users. This discrepancy can lead to misalignment between the users' mental models and their expectations of the application's behavior~\cite{subramonyam2023bridging}. Additionally, teams frequently used the web to search for and retrieve example content during prototyping. When this content is sourced from public or third-party repositories, the legal rights to use it in training AI systems must be carefully considered to avoid copyright infringement. Although this issue was influenced by our study design, future work should explore strategies for \textbf{responsibly using content}, such as developing guidelines for ethically sourcing and utilizing data, ensuring compliance with copyright laws, and implementing robust procedures for obtaining necessary permissions. Finally, designing prompts around specific examples or \textbf{narrowly defined content can lead to overfitting}, where the model struggles to perform well in new or unfamiliar contexts.

Third, by observing role-specific behavior in the study sessions, we highlight the potential for new boundaries and barriers to collaboration in the design process when teams rely heavily on generative AI tools. The widespread use of chat-based interfaces may unintentionally restrict opportunities for innovation in designing more intuitive or creative control mechanisms for end-users. Moreover, teams can become siloed in their approach, focusing primarily on optimizing model outputs rather than holistically exploring the broader design space of the application. While we observed teams employing both convergent and divergent thinking during the design process, further research is needed to explore how teams map and address a wider range of use cases. Recent work, such as Farsight \cite{wang2024farsight}, has introduced methods for designers to identify potential harms during prototyping. Additional research is required to investigate how prototypes can be aligned with considerations around harm and bias. In addition, current frameworks, such as LMSys \cite{lmarena_ai}, primarily capture overall model output satisfaction. However, future directions should emphasize gathering feedback at the interface level, focusing on interaction patterns and user experience beyond the generated model outputs.

\subsection{Limitations}
Given the novelty of generative AI in software products and constraints to gathering process-related data in real-world teams, we focused on a controlled design study with a pre-determined application context, i.e., marketing content creation. While this setup allowed us to glean insights into prompt prototyping and collaboration, it may not be representative of real-world teams. The complexity and variability inherent in different projects, industries, and team structures are likely to influence the process and outcomes of generative AI design in ways not fully explored in our study. Second, we did not include domain experts, in our case, marketing content creators, copyright specialists, etc. While our study is representative of typical roles in software product teams, including individuals with specialized expertise could have influenced how teams curated and interpreted marketing content, leading to different outcomes. Hence, we largely focus our analysis and discussion on the process we observed while cautiously reporting potential pitfalls to content-centric prototyping. Additionally, our focus on a specific application area means that potential pitfalls and limitations identified in our study may not comprehensively cover the spectrum of challenges faced by teams working across various domains and with different types of content and user needs.

%% file: 06_conclusion.tex
\section{Conclusion}
Generative AI models, as a design material, offer dynamic capabilities but also constrain software teams working with pre-trained models. In this study, we explored how collaborative teams prototype generative AI applications through prompt engineering, requiring alignment with human-centered values while addressing diverse user needs. Our findings revealed a shift toward content-centric prototyping, where design and computation are intertwined, and prompts become essential tools for shaping both AI behavior and user experience. However, challenges arose, such as implicit assumptions about user behaviors, overfitting models to narrow content, and the sensitivity of generative AI to prompt variations. These issues highlight the risks of exclusion and misalignment between system design and user expectations. We also observed the ethical concerns of using public content for prototyping, emphasizing the need for guidelines. The shifts in prototyping practices call for a broader exploration of design space and new evaluation frameworks that balance both technical feasibility and user experience in generative AI applications.

\begin{acks}
    We thank the reviewers for their feedback on the paper. We also thank our study participants for their time and input. We are grateful to Emily Reif and Prateek Jain for their thoughtful feedback on the paper. This work is supported by a generous gift from Google.
\end{acks}

%% file: 07_appendix.tex
\appendix
\onecolumn
\section{Appendix: Example prompts crafted by participants in the study}

\begin{figure*}[ht!]
  \centering
\includegraphics[width=0.8\textwidth]{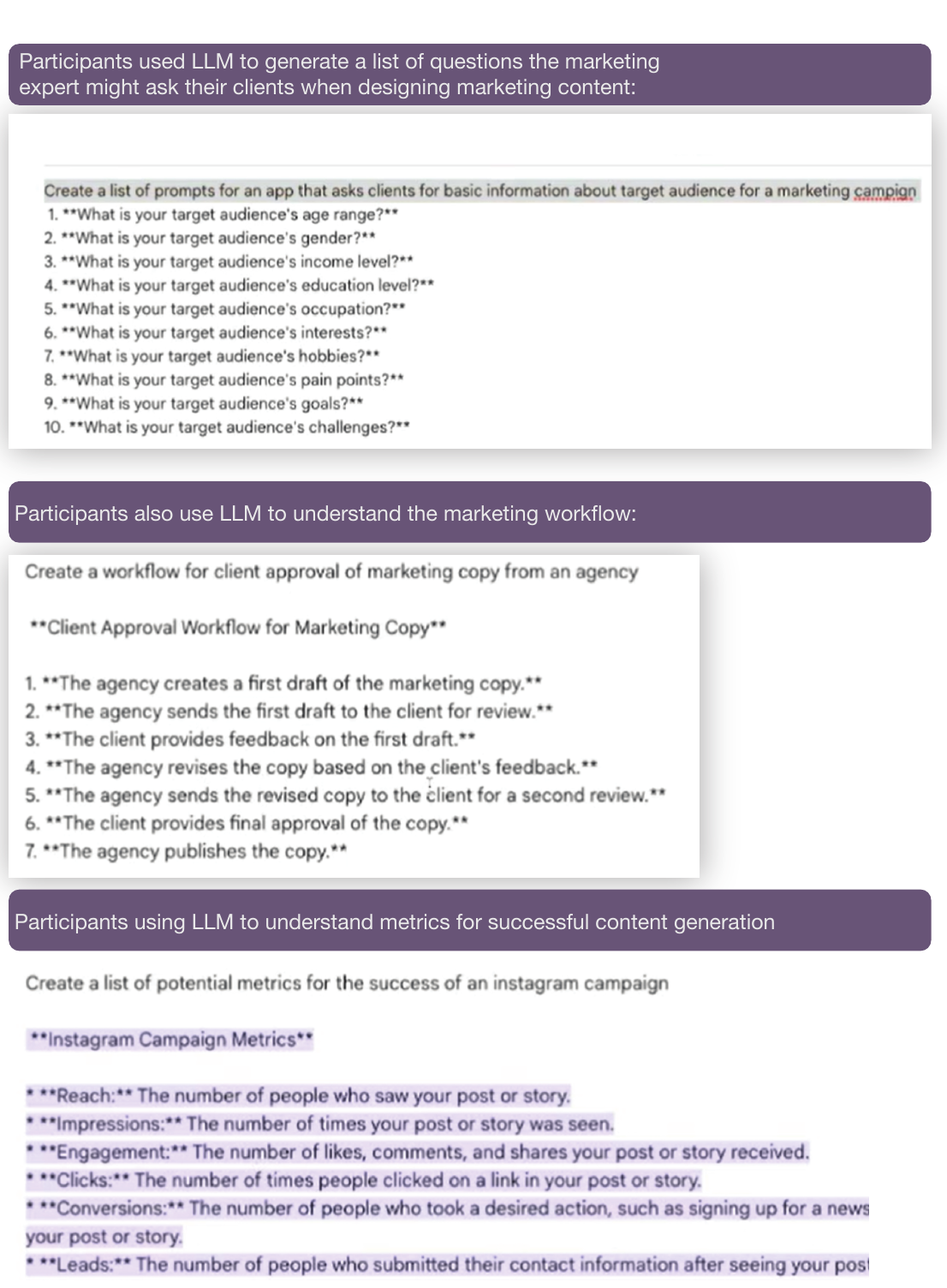}
  \caption{ Example prompts used for need finding using LLM during the collaborative prototyping process. 
}
\label{fig:prompt1}
\end{figure*}

\begin{figure*}[t!]
  \centering
\includegraphics[width=0.85\textwidth]{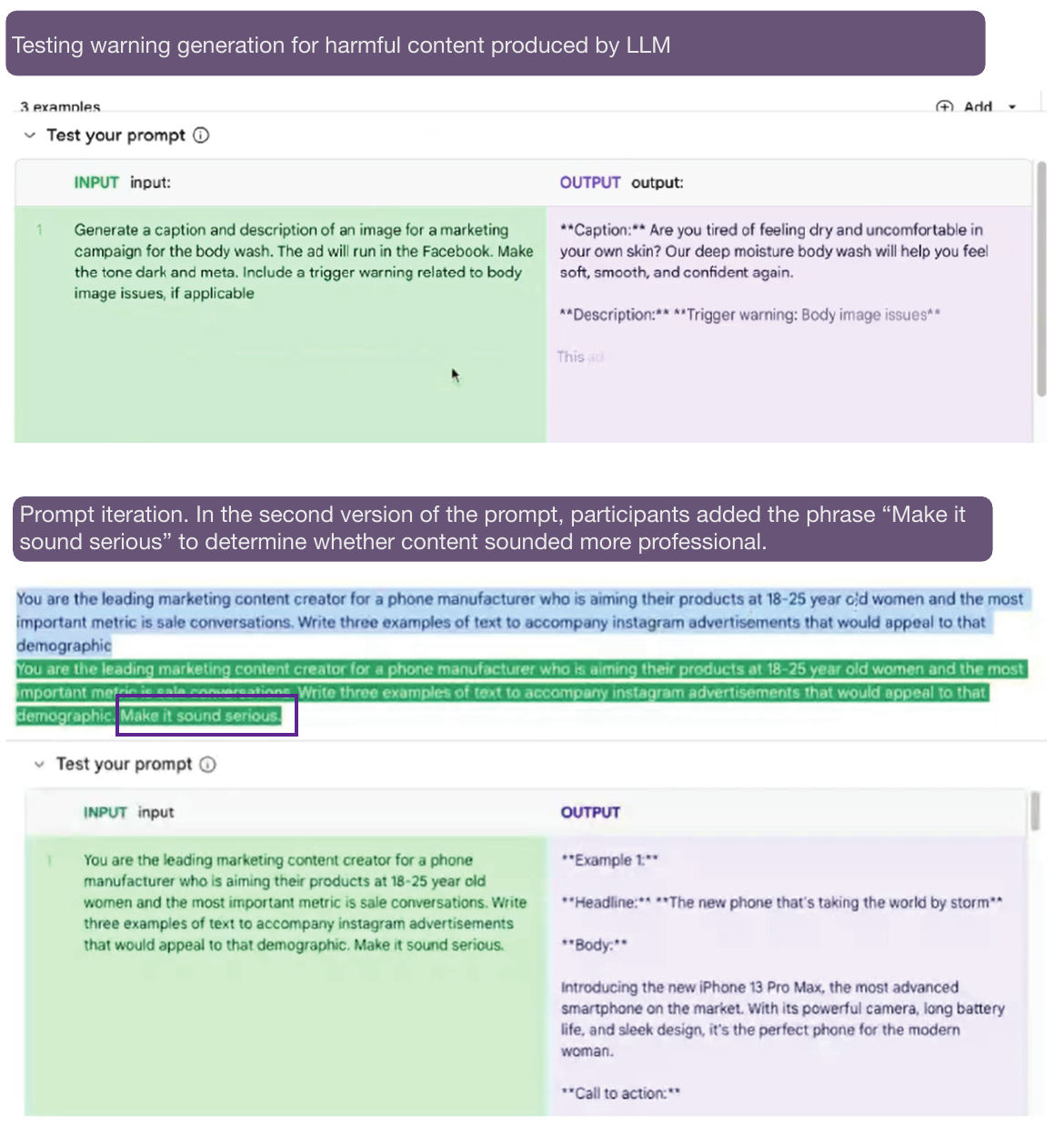}
  \caption{ Iterative prompt authoring by varying prompt phrases and adding warning indicators. 
}
\label{fig:prompts2}
\end{figure*}

\begin{figure*}[t!]
  \centering
\includegraphics[width=0.7\textwidth]{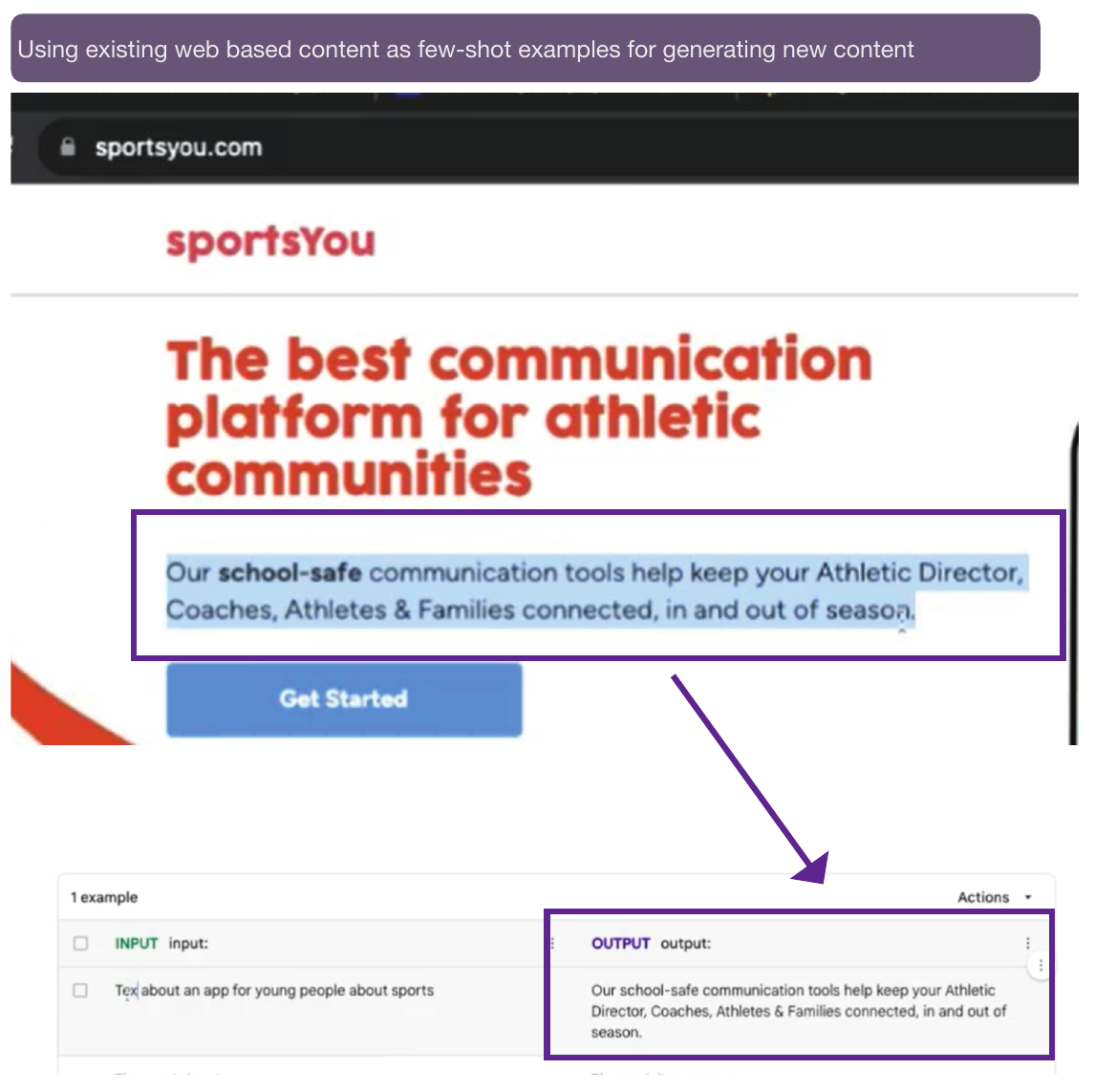}
  \caption{ During prototyping, participants would import content from existing websites as examples for content generation.  
}
\label{fig:prompts3}
\end{figure*}

\begin{figure*}[t!]
  \centering
\includegraphics[width=0.7\textwidth]{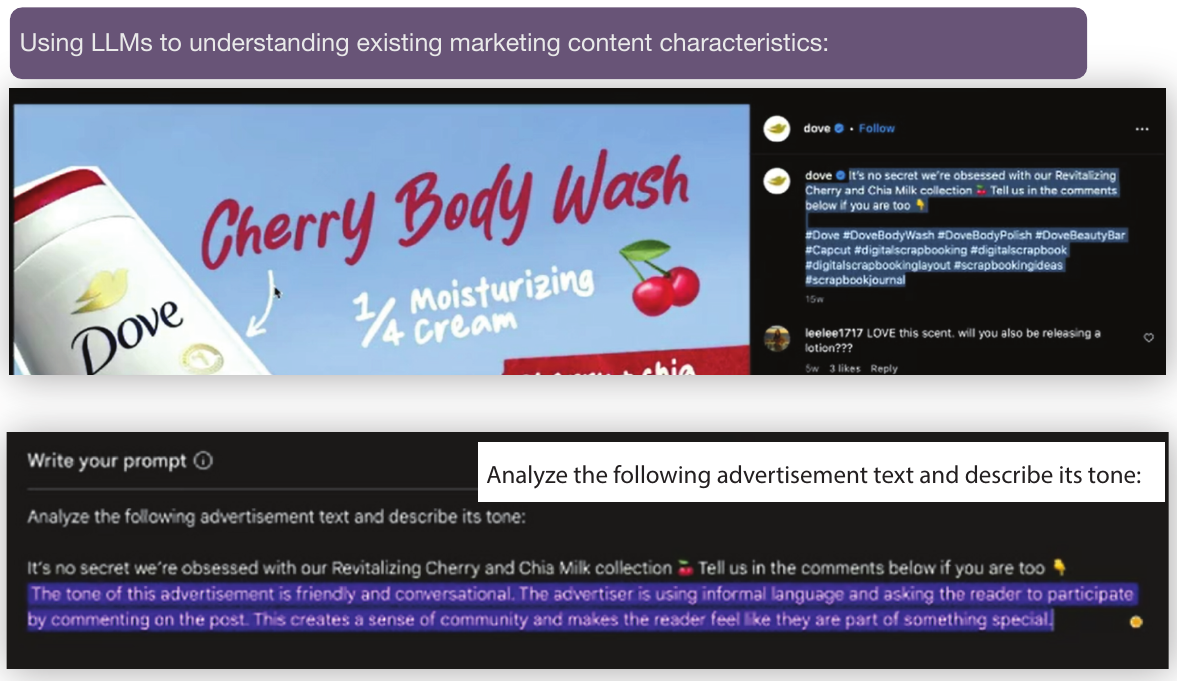}
  \caption{ Participants using LLM to understand specific characteristics of existing marketing content.  
}
\label{fig:prompts3}
\end{figure*}

\begin{figure*}[t!]
  \centering
\includegraphics[width=0.85\textwidth]{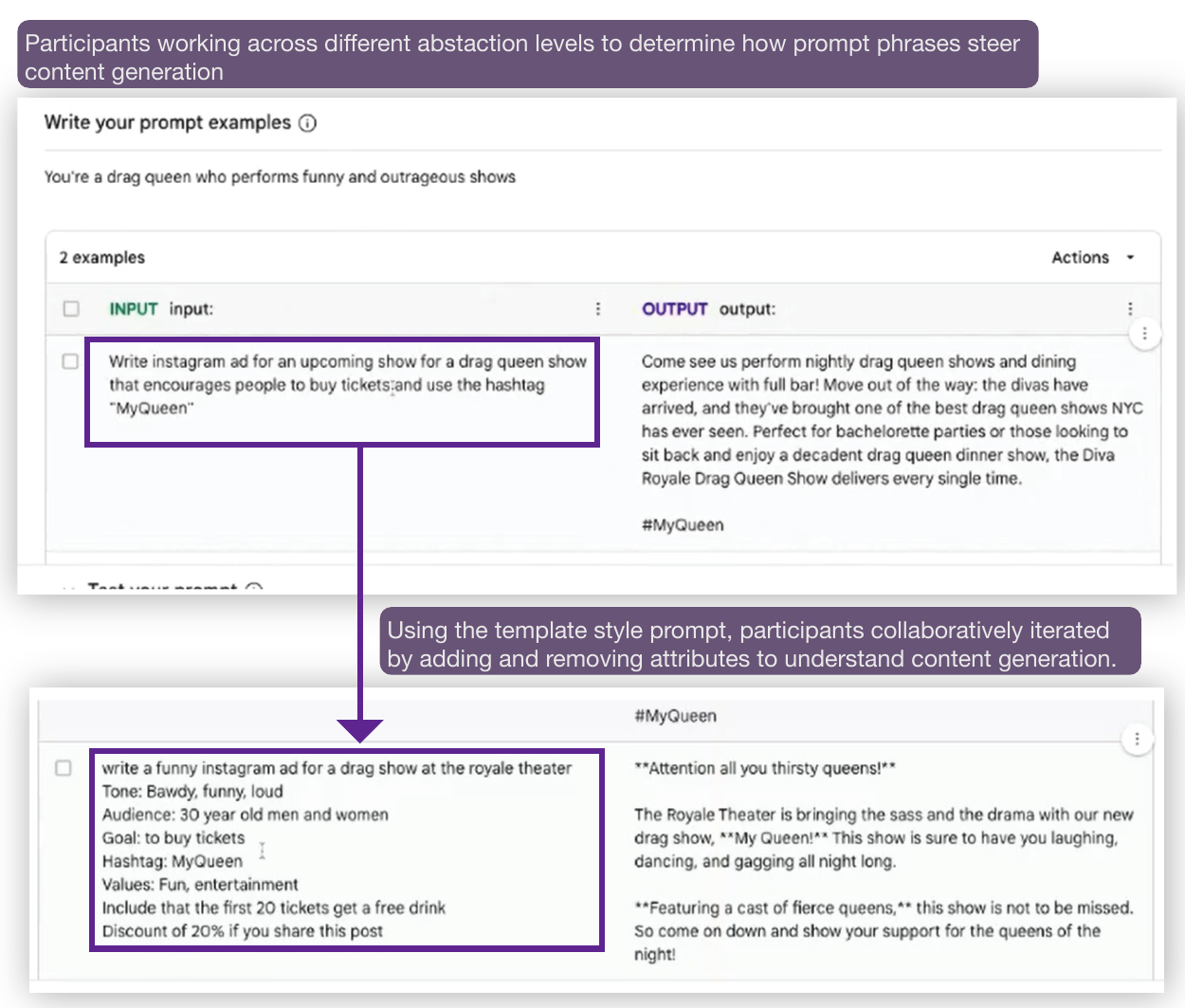}
  \caption{ Prototyping prompts across different abstractions and collaboratively understanding how different input attributes lead to different content.  
}
\label{fig:prompts4}
\end{figure*}

\begin{figure*}[t!]
  \centering
\includegraphics[width=0.85\textwidth]{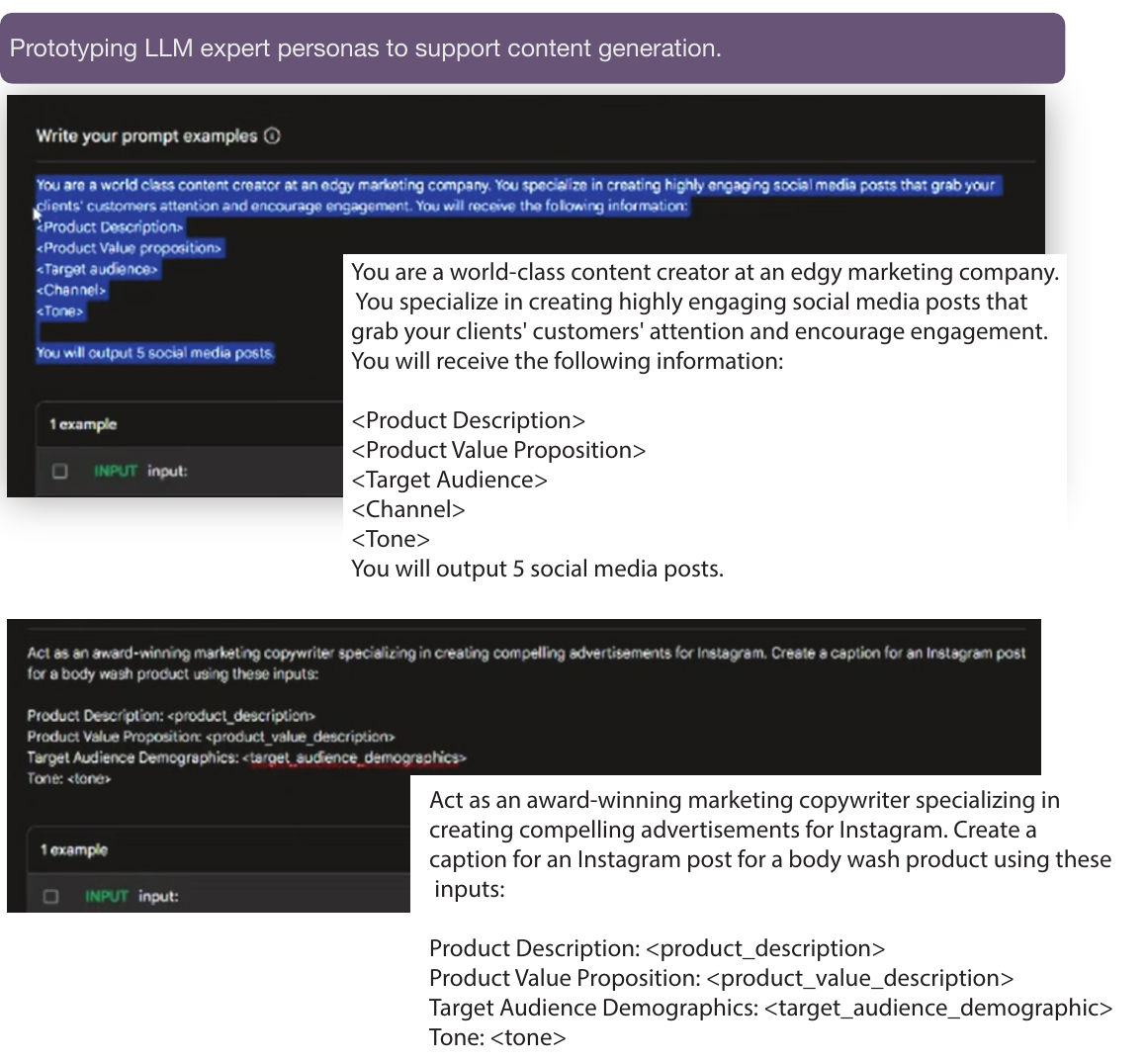}
  \caption{ Prototyping LLM expert personas for content generation. 
}
\label{fig:prompts4}
\end{figure*}